\documentclass[review,12pt]{elsarticle}
\usepackage{hyperref}
\usepackage{amsmath}
\usepackage{graphicx}
\usepackage{float}
\usepackage{subfigure}
\usepackage{array,booktabs,arydshln,xcolor}
\usepackage{epstopdf}
\usepackage{textcomp}
\usepackage{ulem}{}

\DeclareMathOperator*{\argmin}{arg\,min}
\journal{Journal of \LaTeX\ Templates}
\begin{document}

\begin{frontmatter}

\title{A 2D hybrid method for interfacial transport of passive scalars}
\tnotetext[mytitlenote]{Fully documented templates are available in the elsarticle package on \href{http://www.ctan.org/tex-archive/macros/latex/contrib/elsarticle}{CTAN}.}




\author[mymainaddress]{Yu Fan}
\author[mymainaddress]{Yujie Zhu}
\author[mysecondaryaddress]{Xiaoliang Li}
\author[mymainaddress]{Xiangyu Hu\corref{mycorrespondingauthor}}
\cortext[mycorrespondingauthor]{Corresponding author}
\author[mymainaddress]{Nikolaus A. Adams}
\address[mymainaddress]{Technical University of Munich Boltzmannstr. 15 D-85748 Garching }
\address[mysecondaryaddress]{Department of Mathematics "Federigo Enriques", Università degli Studi di Milano,
	Via Cesare Saldini, 50, 20133, Milan, Italy}
\begin{abstract}
A hybrid Eulerian-Lagrangian method is proposed to simulate passive scalar transport on arbitrary shape interface.
In this method, interface deformation is tracked by an Eulerian method while the transport of the passive scalar on the material interface is solved by a single-layer Lagrangian particle method.
To avoid particle clustering, a novel remeshing approach is proposed. This remeshing method can resample particles, adjust the position of particles by a relaxation process, and transfer mass from  pre-existing particles to resampled particles via a redistribution process, which preserves mass both globally and locally.
Computational costs are controlled by an adaptive remeshing strategy.
Accuracy is assessed by a series of test cases.

\end{abstract}


\end{frontmatter}


\section{Introduction}

The presence of surfactants on interfaces in multi-phase flow can alter the surface tension, interface shapes and topology \cite{chu2019review}, and affect the relevant engineering processes \cite{chowdhury2022surfactant,chowdhury2022comprehensive} .
Numerical simulation is one of the most important ways to understand surfactant dynamics.
In simulations, usually, surfactants are modeled as passive scalars.
The computation of the transport of passive scalars is a prerequisite for further study of surfactant dynamics.
Currently, there are several numerical approaches to compute passive scalars: Eulerian, Lagrangian, and hybrid. However, these approaches encounter difficulties with respect to the accuracy, conservation and efficiency.

In the Eulerian approach, the passive scalar on the interface is usually represented on a grid \cite{adalsteinsson2003transport, xu2003eulerian, xu2006level, SCHRANNER2016653, teigen2011diffuse,olshanskii2014stabilized}, which causes a mismatch between the grid data and the off-grid interface, and eventually violates mass conservation.
To cope with this difficulty, a global correction was introduced \cite{SCHRANNER2016653}, 
but still failed to satisfy mass conservation locally.
Another approach proposed for insoluble surfactants based on a volume-of-fluid method 
tracks mass and volume separately, which enforces the conservation of former \cite{james2004surfactant}.
However, for large deformation or long-time simulation, 
the consistency error between mass and volume can increase considerably.

Contrary to Eulerian methods, Lagrangian methods maintain mass conservation inherently.
A typical example is Smoothed Particle Hydrodynamics (SPH), 
which has been used to study surfactant dynamics on bubble surfaces \cite{adami2010conservative}.
The interface was represented by multiple layers of particles 
so that a 2D problem is transformed into 3D. 
Since a 3D problem involves more neighbor connections 
when computing physical variables, 
it leads to extra computational costs in run-time search of the neighbor connections.
To save costs, a codimension-1 SPH method, based on differential geometry, 
was proposed for surfactant dynamics on thin film and bubbles \cite{wang2021thin}.
However, as the particles are added by reseeding 
when the surface is enlarged, mass is not conserved.

In contrast to either pure Eulerian or Lagrangian approaches, many researchers propose hybrid approaches, 
which take advantage of both properties. 
While the interface is tracked by an Eulerian method, 
Lagrangian surface meshes are used to solve the advection-diffusion equation of the passive scalar \cite{ceniceros2003effects,lai2008immersed,lai2010numerical,de20153d}. 
The difficulty is that, 
when the surface mesh is deformed by the flow, 
its mesh quality can not be guaranteed.
To avoid degeneration of mesh quality, 
an artificial tangential velocity \cite{hou1994removing} is introduced in Refs. \cite{ceniceros2003effects, lai2010numerical}.
Other similar methods proposed to solve this problem 
used either a map between the initial parameter of the curve and the arc-length parameter of the current interface \cite{lai2008immersed}, 
or a surface mesh optimization method \cite{botsch2004remeshing}.
Although these methods can alleviate the problem of mesh quality, this geometric correction is quite complicated.
An additional problem of these hybrid methods is that, 
to solve the transport equation on the surface, 
connectivity is necessary for all these methods.

In this work, 
we investigate an alternative hybrid approach, 
by using the mesh-free particle Lagrangian formulation 
to discretize the transport equation of the passive scalar on a co-dimension 1 interface while tracking the interface motion by the level-set method.
This discretization is on a single layer of particles, 
by which the mass conservation is satisfied automatically.
Since the particles are only distributed on the co-dimension 1 interface, 
2D problems are reduced to 1D,
every particle has less neighboring connections,
and computational costs are saved significantly.
The method does not require explicit geometry/topology information and higher-order derivatives, which simplifies implementation and improves robustness.
When the interface experiences strong stretching or compression,
a generalized remeshing method is proposed based on classical remeshing \cite{cottet2000vortex,rossinelli2008vortex}.
This generalized remeshing method can achieve 
arbitrary order consistency and 
the mass conservation holds automatically as the zeroth order consistency.
Besides, this method is generalized for multi-region problems based on the regional level-set method \cite{zheng2009simulation,pan2018high}.
The availability of particles can also be used to improve the accuracy of the regional level-set method without taking too much extra memory.

The remainder of this paper is organized as follows:
Hybrid method and its generalized version for multi-region problems are detailed in Section 2.
Section 3 comprehensively validates the accuracy, 
efficiency and conservation through several benchmark tests.
\section{The hybrid method}
\begin{figure}[tb!]
	\centering
    \includegraphics[width=1.0\textwidth]{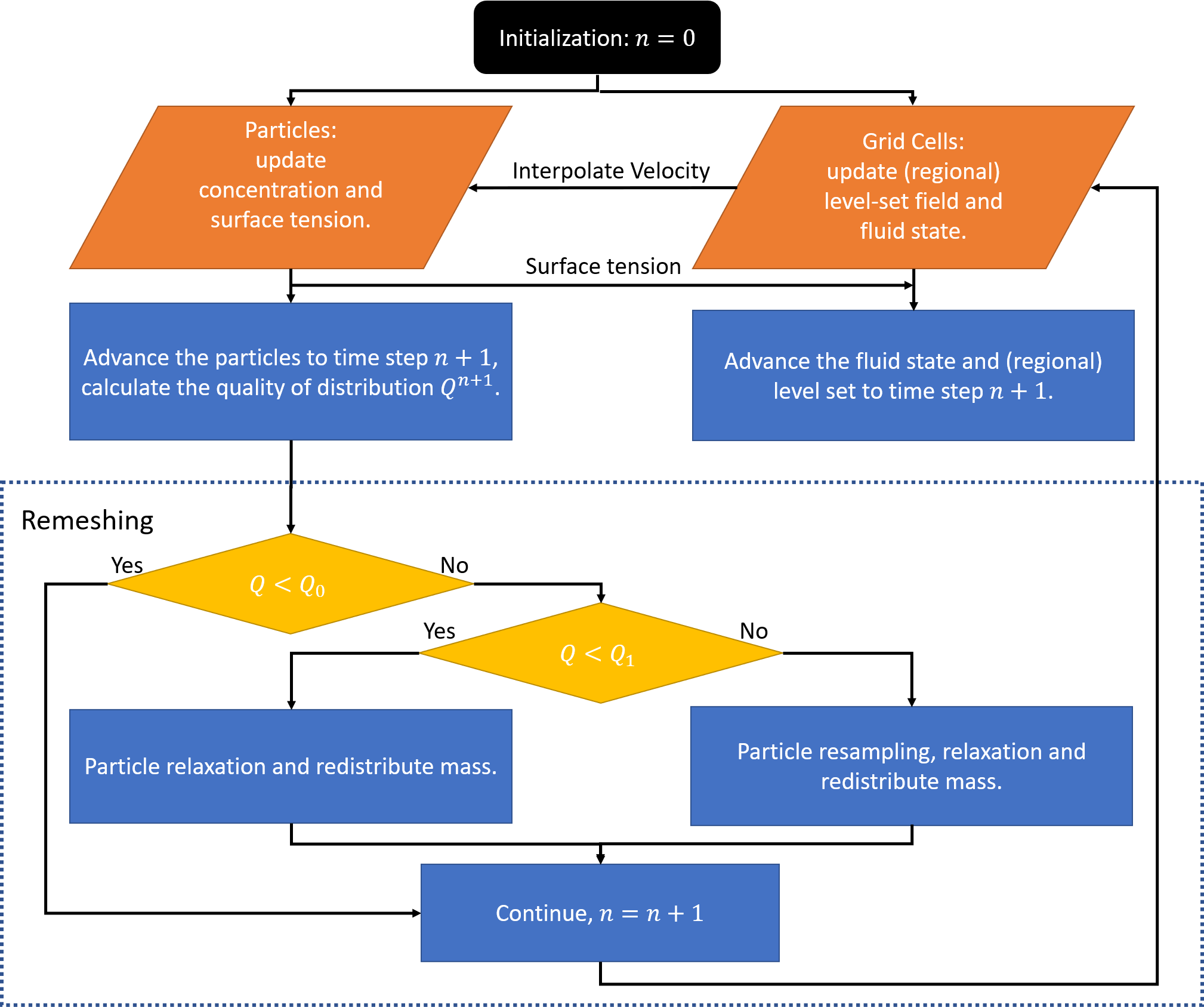}
	\caption{Flow chart for the hybrid method.}
    \label{flowchart}
\end{figure}

In this section, details of the hybrid Eulerian-Lagrangian method are presented.
We adopt the conservative interface method \cite{hu2006conservative} to track the interface deformation, 
while a Lagrangian particle method is
used to compute the evolution of passive scalar on the interface, see Fig. \ref{flowchart}.
Particles are generated on the interface based on the level-set field 
and regularized by a relaxation process.
At each time step, the position of each particle is updated by the velocity interpolated 
from the neighboring cell centers.
If the particles are unevenly distributed after advection,
a generalized remeshing method will modify both the position of particles 
and the value of passive scalars.
To reduce the expense, an adaptive strategy is introduced to decrease the frequency of remeshing.

\subsection{The Eulerian component}
	On the Eulerian side, we use a level set function $\phi(x,y,t)$ to represent the interface \cite{osher2001level}.
	The level set function has the property $|\nabla\phi| = 1$ as signed distance,
	and identifies the inside region from the outside by the signs:
	$\phi<0$ corresponds to the inside, $\phi>0$ the outside,
	and $\phi=0$ the interface.
	The level set function can be updated by solving the advection equation
	$\frac{\partial \phi}{\partial t} + u\phi_x+v\phi_y = 0$
	with reinitialization steps \cite{sussman1994level,fu2017single}.
    The velocity field is either given with analytical function for simple cases 
    or obtained from the solution of two-phase N-S equations \cite{hu2006conservative}.

\subsection{The Lagrangian component}
We use Lagrangian particles to solve the transport equation of the passive scalar $\frac{\partial C}{\partial t} + \vec{u} \cdot \nabla_s C=0$, where $\nabla_s$ is the gradient on the surface. 
The mass of every particle is initialized as $m_p = C_pV_p$,
where the concentration $C$ is assigned by a given initial condition.
The concentration field can be approximated by a
 weighted average of the mass of the particles within the cut-off radius.
\begin{equation}
C_p = \Sigma_q{W_{pq}m_q}.
\end{equation}

There are three main steps to solve the transport equation: particle initialization, particle advection and remeshing.
The first step is to initialize the particles. We generate the particles at the centers of cut cells, which can be identified by the signs of level set values at 9 points ($8$ points on cell boundary and one at cell center).
Then, the particles are pushed to the interface with a distance $|\phi|$ in the direction $-sgn(\phi)\vec{n}$.
However, this distribution of particles is non-uniform, which will cause a zeroth-order inconsistency.
\begin{figure}[tb!]
	\centering 
	\includegraphics[width=0.5\textwidth]{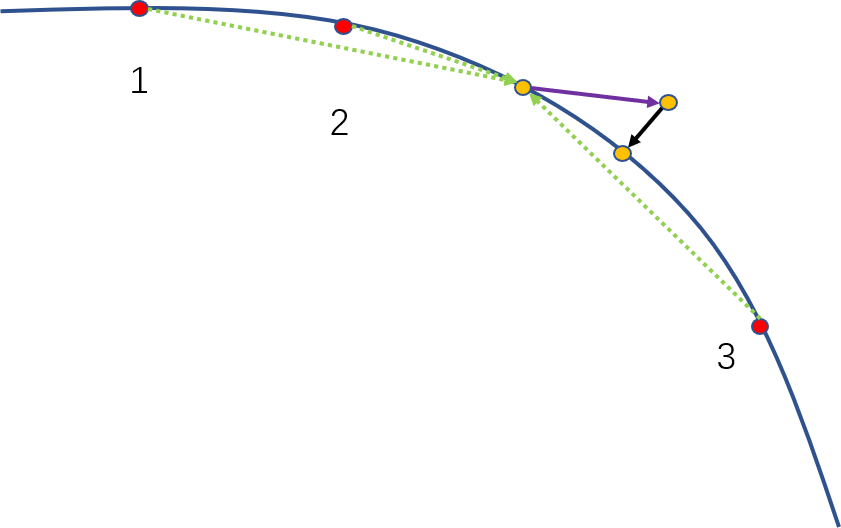}
	\caption{A sketch for the relaxation algorithm. The red and yellow particles are the neighboring particles and the considered particle respectively. The position is adjusted (purple arrow) by the resultant force of the repulsive force (green dashed arrow) between every pair of particles, and then confine the particles to the interface (black arrow).}
	\label{relaxation}
\end{figure}
To improve the consistency, a relaxation method proposed by Fu et al. \cite{fu2019isotropic}
and simplified by Zhu et.al \cite{zhu2021cad} for SPH particle generation,
is adopted here to assure a uniform distribution of particles along the interface, see Fig. \ref{relaxation}.
The second step is to advect the particles with the velocity interpolated from the level set field by Eq. (\ref{advection}).
\begin{equation}
\frac{d\vec{x_p}}{dt}=\vec{u}.
\label{advection}
\end{equation}
The final step is to remesh particles for avoiding particle clustering during advection. 
The remeshing process consists three sub-steps: resampling (optional), relaxation, and redistribution.

\begin{figure}[tb!]
	\centering 
	\includegraphics[width=0.5\textwidth]{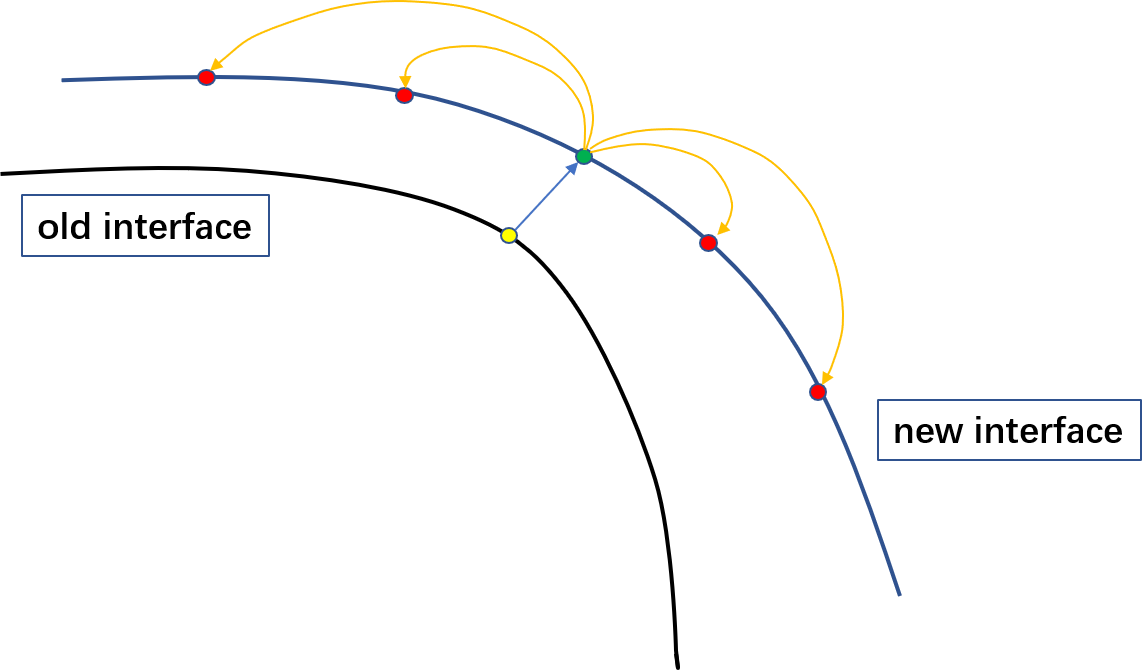}
	\caption{Redistribution of the mass from old cloud (green points) to new cloud (red points). The yellow point and black curve is the particle and the interface at current time step, while the green point and blue curve is the particle and interface at next time step.}
	\label{redis}
\end{figure}

The resampling enables adjusting the number of particles, 
which can be very helpful to avoid void regions when the interface is extremely stretched. 
However, after resampling a uniform distribution of particles often requires many relaxation iterations.
To reduce the frequency of remeshing, 
we introduce a criterion based on a measure of the particle distribution, 
\begin{equation}
	Q = \frac{\max\{|v_i-\bar{v}|\}}{\bar{v}}
	\label{quality}
\end{equation}
where $\bar{v}=\sum_p v_p/n$ is the average volume of particles and $n$ is the number of particles, $v_p=\frac{1}{\Sigma_q{W_{pq}}}$ is the smoothed volume \cite{hu2007incompressible} of the particle $p$, and $W_{pq}$ is the 5th-order Wendland kernel function $W(|x_p-x_q|,h)$ with smoothed length $h$.
Uniform distributions are indicated by $Q \ll 1$. We can measure the distribution quality every $N$ time steps and in the following we set $N=30$. Accordingly, we use the following remeshing criterion to control $Q$:

\begin{itemize}
\item $Q<Q_0$, we remesh particles. In the following we use $Q_0 = 0.01$.

\item $Q_1<Q$, we resample particles followed by relaxation iterations until $Q<Q_0$.
Finally, we redistribute the mass from old particles to the new ones.
In the following we set $Q_1 = 0.03$.

\item $Q_0 \leq Q \leq Q_1$, 
we do not resample particles but do $M$ times relaxation iterations. To save costs, $M$ should be much smaller than the number of iterations when $Q>Q_1$.
In the following we use $M=30$.  
Finally, we redistribute the mass from pre-existing particles to the resampled ones.
\end{itemize}
The redistribution process transfers the passive scalar from the old particles to the new ones, 
see Fig. \ref{redis}.
The redistribution method is generalized from the traditional particle remeshing method \cite{cottet2000vortex}.
The key idea is to use a new set of particles to represent the old field of passive scalar represented by the old ones. 
We can define the error of concentration between the two set of particles as
\begin{equation}
	\epsilon_c(x) = C^{old}(x)-C^{new}(x),
	\label{deferror}
\end{equation}
where $C^{old}$ and $C^{new}$ are the concentrations 
represented by the old and new particle sets, respectively. 
The redistribution process can be expressed as
\begin{equation}
	m_q = \Sigma_p{m_p\beta_{pq}},
	\label{defredistribution}
\end{equation}
where $\left\{x_p	\right\}$ is the old set, 
$\left\{x_q	\right\}$ is the new one, 
and the subscript $m$ means the particle mass.
In order to preserve the accuracy, 
we can look for a proper set of coefficients $\beta_{pq}$ 
in Eq. (\ref{defredistribution}) to minimize the error defined 
in Eq. (\ref{deferror}) to a given order $O(h^p)$. 
Here we let $p=4$ to balance the cost and accuracy.

The $C^{old}(x)$ and $C^{new}(x)$ in Eq. (\ref{deferror}) can be further written as
\begin{equation}
	C^{old} = \Sigma_p{W(|x-x_p|,h)m_p}, \quad
	C^{new} = \Sigma_q{W(|x-x_q|,h)m_q}.
\end{equation}
Thus the error becomes
	\begin{align}
	\label{error}
	\epsilon_c(x) &= C^{old}(x)-C^{new}(x) \\
		  	 &=\Sigma_p{W(|x-x_p|,h)m_p} - \Sigma_q{W(|x-x_q|,h)m_q} \\
		  	 &=\Sigma_p{W(|x-x_p|,h)m_p} - \Sigma_q{W(|x-x_q|,h)\Sigma_p{m_p\beta_{pq}}}\\
		  	 &=\Sigma_p{m_p[W(|x-x_p|,h)- \Sigma_q{W(|x-x_q|,h)\beta_{pq}}]}.\label{finalerror}
	\end{align}
The function $W(|x-x_q|,h)$ can be approximated by 
a Taylor expansion at $(x-x_p)$ to third order as
\begin{equation}
	\label{Taylor}
	W(|x-x_q|,h) = W((|x-x_p|,h)+\Sigma_{k=1}^3{W_y^{(k)}|_{y=x_p}\frac{(x_q-x_p)^k}{k!}}+{\cal O}((x_p-x_q)^4).
\end{equation}
We can minimize the error by eliminating the first 4 terms of
 $(x_q-x_p)^k$ with suitable weights $\beta_{pq}$ from
	\begin{equation}
    \label{linearsystem}
	\left (
	\begin{array}{cccc}
	1 & 1 & 1 & 1\\
	x_p-x_1 & x_p-x_2 & x_p-x_3 & x_p-x_4 \\
	(x_p-x_1)^2 & (x_p-x_2)^2 & (x_p-x_3)^2 & (x_p-x_4)^2 \\
	(x_p-x_1)^3 & (x_p-x_2)^3 & (x_p-x_3)^3 & (x_p-x_4)^3
	\end{array}
	\right)
	\left (
	\begin{array}{c}
	\beta_1 \\
	\beta_2 \\
	\beta_3 \\
	\beta_4 \\
	\end{array}
	\right)
	=
	\left (
	\begin{array}{c}
	1 \\
	0 \\
	0 \\
	0 \\
	\end{array}
	\right)
	\end{equation}
This coefficient matrix is a transformed Vandermonde matrix, 
such that the solution exists and is unique:
	\begin{align}
	\beta_0 &= \frac{x_4}{x_4-x_1}\cdot \frac{x_3}{x_3-x_1}\cdot \frac{x_2}{x_2-x_1}\\
	\beta_1 &= \frac{x_4}{x_4-x_2}\cdot \frac{x_3}{x_3-x_2}\cdot \frac{x_1}{x_1-x_2}\\
	\beta_2 &= \frac{x_4}{x_4-x_3}\cdot \frac{x_2}{x_2-x_3}\cdot \frac{x_1}{x_1-x_3}\\
	\beta_3 &= \frac{x_3}{x_3-x_4}\cdot \frac{x_2}{x_2-x_4}\cdot \frac{x_1}{x_1-x_4}	
	\end{align}
Apparently, this method can be generalized to arbitrary order if 
there are enough neighboring particles in the compact support of each particle.

\subsection{Extension to multi-region problems}
The multi-region problem can be solved 
with a multiple level-set method \cite{merriman1994motion}, 
and combined with the present method in a straightforward way. 
However, with increasing number of regions,
the multiple level-set method demands larger memory size 
because each region is represented by a separated level-set field. 
This problem is avoided by a regional level-set method
\cite{zheng2009simulation,pan2018high}, 
in which only a pair of data is stored in the memory, 
i.e. an unsigned distance function and a region index (also called as color index). However, to implement the regional level-set method, some modifications are necessary. Before introducing the modifications, we review the mechanism and some related basic concepts of the regional level-set method.

Similar to classical level-set method, each region is still tracked by its own level-set field, but this level-set field, called local level-set field, is constructed from the regional level-set field.
After advancing the local level-set field, the regional level-set field can be updated by the new local level-set field.
With assignment of all indicator colors, the updating of the regional level-set field is done. This process can be abstracted as a reconstruction mapping, which maps all local level-set fields to a regional level-set field.

We denote the local level-set function of color index $r$
as $\phi^r(x,y)$,
with positive value inside and negative outside,
and the regional level-set function as $\psi=(\phi(x,y),r(x,y))$,
where the function $\phi$ here is the unsigned distance function from the cell center to the nearest interface, and $r$ is its indicator color.
Because the local level-set fields will not be stored in the memory, it needs to be constructed by the construction mapping based on the regional level-set field.
The construction mapping from regional level-set field to local level-set field with color $\chi$ can be represented as
	\begin{equation}
	\phi^\chi = \mathcal{C}^\chi(\phi, r) =\bigg\{
	\begin{aligned}
	\phi, \quad r = \chi \\
	-\phi, \quad r \neq \chi
	\end{aligned},
	\end{equation}
where $\mathcal{C}^\chi$ is the construction mapping from regional level-set field to local level-set field indexed $\chi$. If $\chi$ is removed, $\mathcal{C}$ is the reconstruction mapping from regional level-set field to all local level sets. 

Because only the regional level-set field is stored in the memory, the regional level-set field needs to be updated from all local level-set fields. 
For cell $C$ whose center is $(x,y)$, if there exists at least one positive local level-set field, the reconstruction mapping can be defined as
	\begin{equation}
	\psi=(\phi(x,y),\chi(x,y)) = \mathcal{R}(\{\phi^r,r \in I^+\}) = (\min_{ r \in I^+}\{\phi^r\}, \argmin_{ r \in I^+}(\phi^r)),
	\label{construction}
	\end{equation}
where $I^+$ is the set of local indices with positive local level-set value.
If all the local level-set values are negative,
we need to choose the one with smallest absolute value, and the reconstruction mapping becomes
	\begin{equation}
	\psi(\phi(x,y),\chi(x,y)) = \mathcal{R}(\{\phi^r,r \in I\}) = (\min_ {r \in I}\{|\phi^r|\}, \argmin_{ r \in I}(\phi^r)),
	\end{equation}
where $I$ is the set of all local indices.
The constructing operator is not one-to-one, which causes an inconsistency $\mathcal{RC}((\phi,\chi)) \neq (\phi,\chi)$, see Fig. \ref{error}. 
This will introduce extra error and lead to an inaccurate concentration field of the passive scalar.
Thus, we introduce some extra treatments to reduce the error.
Here, we consider the application of present method with 
the $2D$ regional level-set method.
\subsubsection{Correction near triple points}
\label{correctionsrl}
\begin{figure}[tb!]
	\centering
	\includegraphics[width=0.5\textwidth]{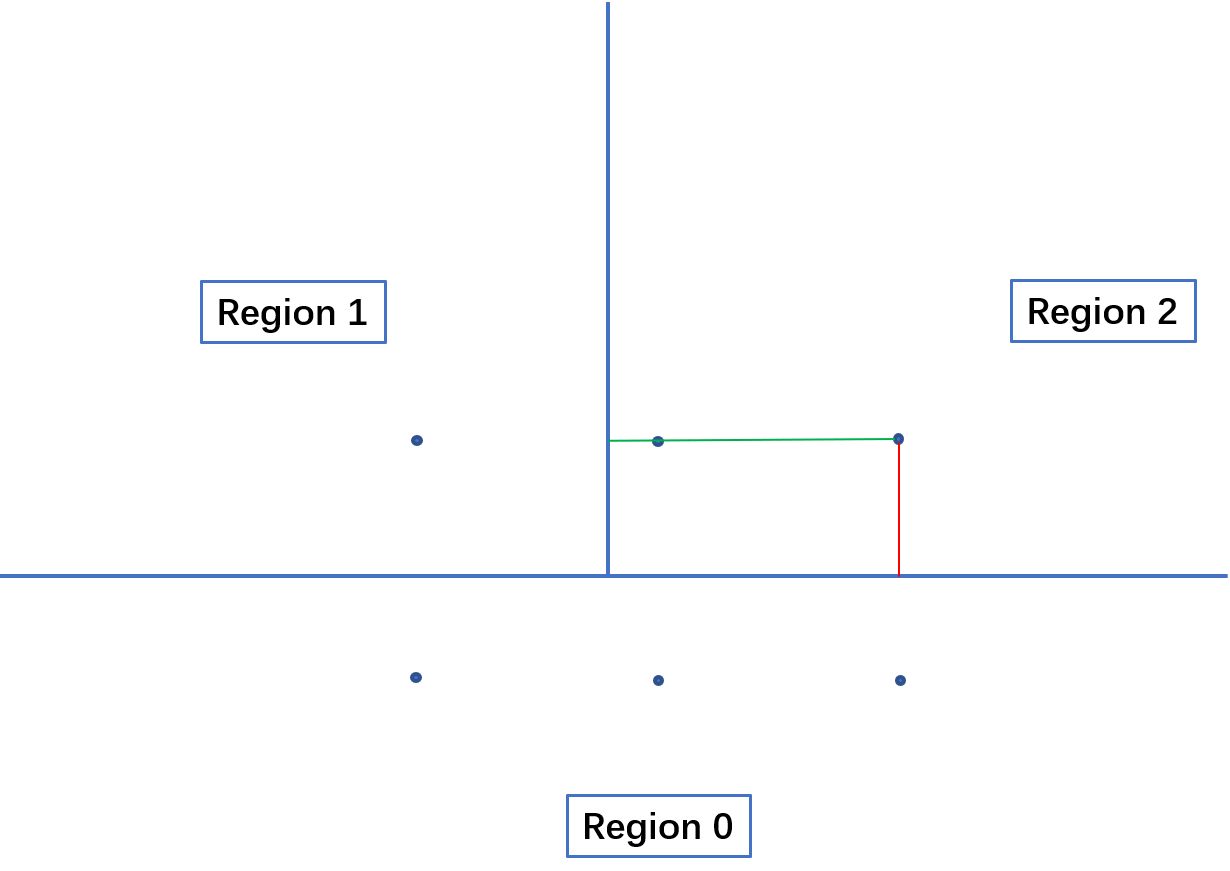}
	\caption{An example of the error introduced by regional level-set method near the triple points.
	For the level-set field of region $1$, the green line is the exact distance while the red line is the inaccurate distance reproduced by the regional level set method.}
\end{figure}
We introduce two techniques to fix this problem: a particle-based correction for the local level-set field and a local mass exchange to smoothing the passive scalar.
Since particles on the interface are readily available, we can use them to improve the local level-set field without requiring additional memory.
Specifically, we utilize particles to correct the distance in $3\times3$ set of neighboring cells around the triple point.
There are two important parts for correcting the local level-set field
, the correction of the sign and of the distance.
\begin{figure}[tb!]
	\centering
	\includegraphics[width=0.6\textwidth]{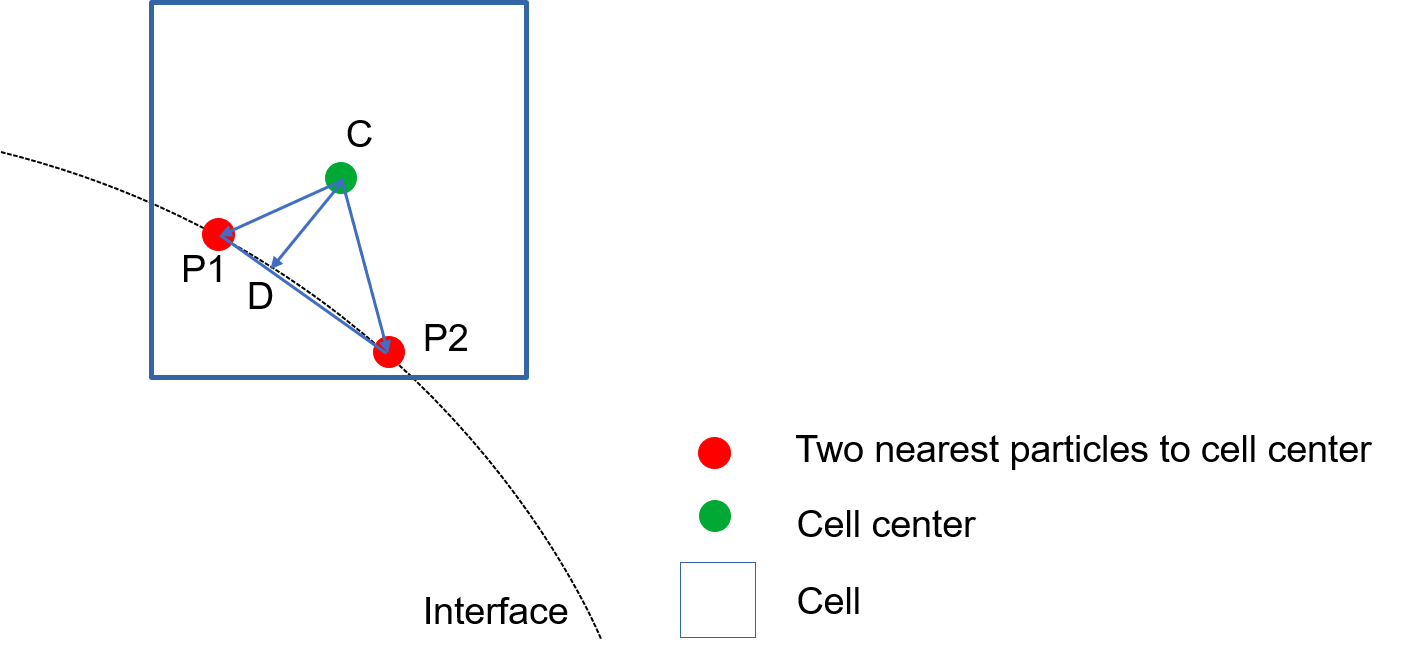}
	\caption{Measuring the distance. Looking for two nearest particles to the cell center $C$, denote them as $P_1$ and $P_2$. Calculate the height of  $\triangle CP_1P_2$ to the edge $P_1P_2$ as the distance.}
	\label{Measuring the distance}
\end{figure}
The simplest method to estimate the distance
is looking for the 2 nearest particles of a cell, see Fig. \ref{Measuring the distance}.
\begin{figure}[tb!]
	\centering
	\includegraphics[width=0.5\textwidth]{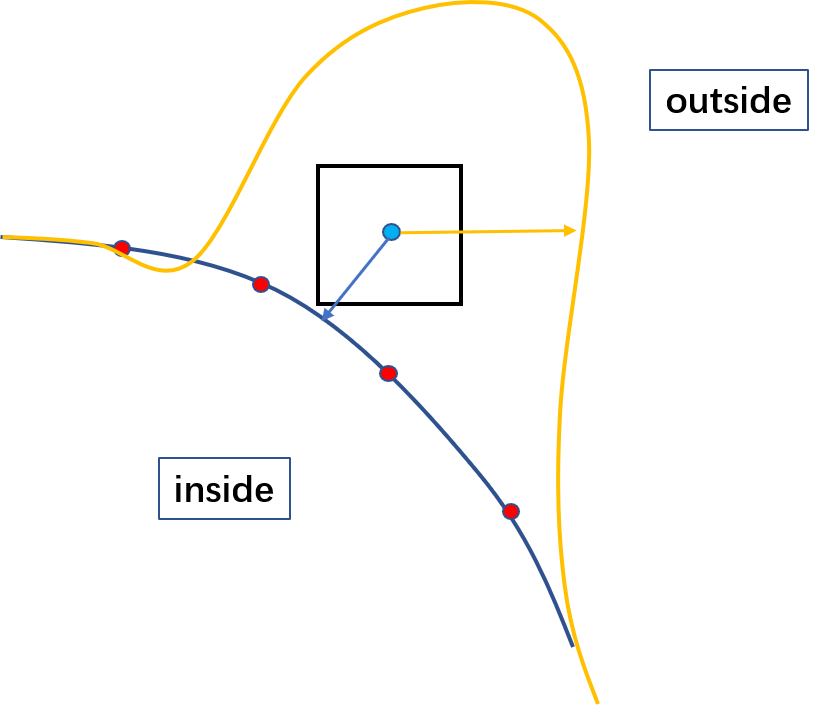}
	\caption{
		Determination of the sign of the level-set function near a triple point.
		Yellow and blue arrows are the normal vectors estimated by the disturbed interface (yellow curve) and particles (red points) respectively.}
	\label{sign}
\end{figure}
On the other hand, a wrong sign is more critical because it may introduce a very large error, which can be about two times as the distance.
However, this can be modified by using particles as well.
Firstly, we estimate the interfacial normal vector from particles, see Fig.
\ref{sign}. Its direction points from the cell center to the interface segment.
We also have a normal vector estimated by the the local level-set field $\vec{n}=\frac{\nabla \phi}{|\nabla \phi|}$.
If the angle between these two normal vectors is larger than 90 degrees, the sign of local level-set is wrong and should be changed.

The second technique is to introduce mass exchange near the triple point to reduce the oscillations of concentration field. 
Because of the limited resolution provided by the particles, the error introduced by the regional level-set method can not be resolved completely by the above correction.
In a $3\Delta x \times 3\Delta y$ neighboring region of the triple point,
we introduce a mass exchange between each pair of distinct particles $p$ and $q$
\begin{equation}
f_{pq} = -(M_p-M_q)W_{pq}\frac{V_p+V_q}{2},
\end{equation}
where $f_{pq}$ is the net amount of mass exchanged between particle $p$ and particle $q$.
$M_p$ and $M_q$ are the mass of the two particles respectively,
$W_{pq}$ is the weight function of the distance between $p$ and $q$,
and $V_p$ and $V_q$ is the volume of the particles respectively.

The corrected mass of particle $p$ is $M_p^*$
\begin{align}
M_p^*
&=M_p-\sum_{q\neq p}{(M_p-M_q)W_{pq}\frac{V_p+V_q}{2}}\\
&=M_p-\sum_q{(M_p-M_q)W_{pq}\frac{V_p+V_q}{2}}\\
&\approx \sum_{q}{M_qW_{pq}V},
\end{align}
in which the last approximation results from assuming that the volume of the particles is nearly equal to the average volume $V$.
This mass exchange makes the mass of every particle in this region be a weighted average of all the other particles in this region.
Obviously, this process maintains the mass conservation. 
Moreover, the measure of influenced neighboring cells (namely the total area) tends to zero as the resolution tends to infinite, thus this mass exchange is consistent.

\section{Test cases}
Different test cases are considered in this section.
The simplest one is translational movement.
The uniform case tests the error from the remeshing process,
and the non-uniform case measures the numerical diffusion.
Furthermore, a droplet in the given simple shear flow without surface tension and in two-phase shear flow with surface tension is considered.
Besides the multi-region case tests the compatibility with regional level set method.

\subsection{Translation of passive scalar distributed on a circle}
\label{chap:translation}
	The transport of passive scalar with uniform and non-uniform initial distribution on a unit circle in a uniform and constant flow field is tested. 	
	Theoretically,
	the concentration of the passive scalar on the interface should remain unchanged.
	
	The computational domain is $[0,1]\times[0,1]$ on the $(x,z)$ plane.
	The circle with radius $r=0.2$ is centered at $(0.3,0.3)$
	in a constant flow $u=0.4/\sqrt{2}$ and $w=0.4/\sqrt{2}$.
	The final time is $t_{final}=1.0$.
	The initial concentrations are $1.0$ and
	$c(\theta)= 2+\cos(\theta)$, where $\theta$ is the angle,
	for the uniform and non-uniform cases, respectively.

	To study the accuracy of remeshing and control for the impact of particle distribution, we conduct particle relaxation at each time step with a sufficiently large number of iterations to achieve a uniform particle distribution. However, in practical use, remeshing is typically only required when particle distribution becomes degenerate, and therefore a much smaller number of iterations is usually sufficient.

\begin{figure}[tb!]
	    \subfigure[Initial Condition]{
			\begin{minipage}[b]{0.45\textwidth}
				\includegraphics[width=1\textwidth]{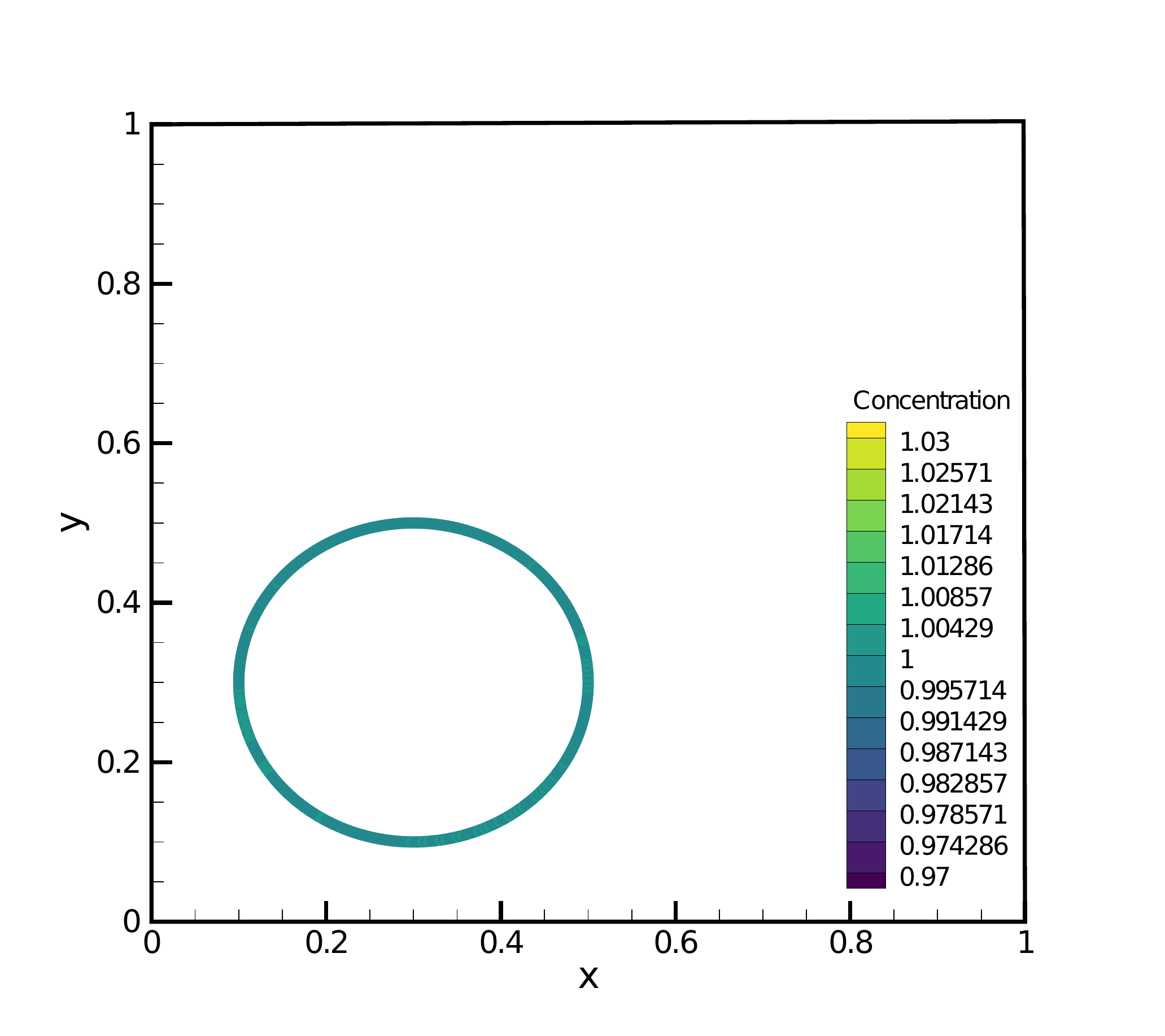}
			\end{minipage}
			\label{fig:TUI256}
		}
    	\subfigure[$t=1.0$]{
    		\begin{minipage}[b]{0.45\textwidth}
   		 	\includegraphics[width=1\textwidth]{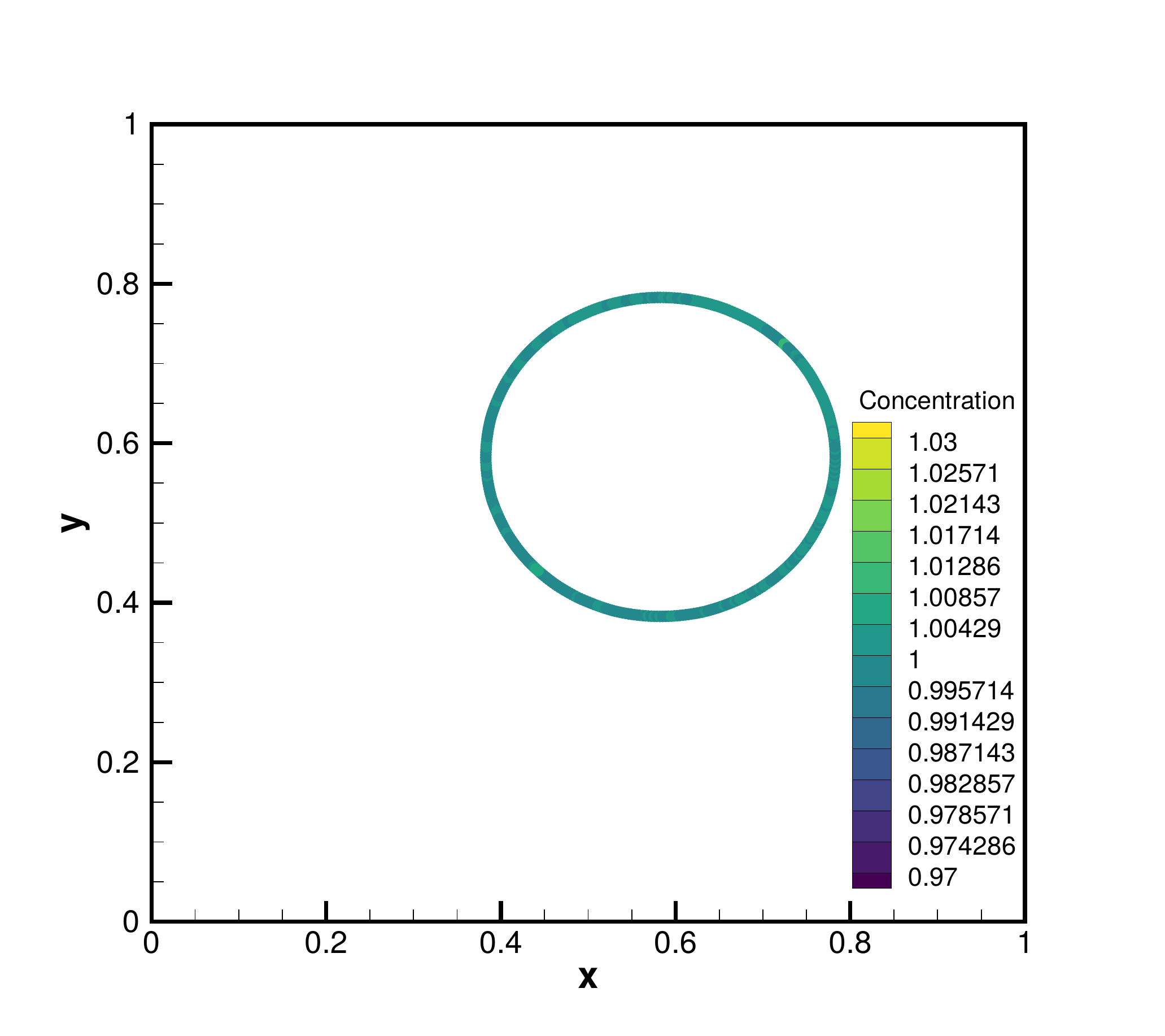}
    		\end{minipage}
			\label{fig:TUF256}
		}
		\caption{Translation case with a uniform initial concentration, $256\times256$ cells}
		\label{TU}
	\end{figure}

	In the first part, the uniform case is tested to verify the zeroth-order consistency of the redistribution method.
	Fig. \ref{TU} and Tab. \ref{uniformtab} show that our redistribution method has good zeroth-order consistency and convergence properties when a sufficiently uniform particle distribution is achieved.

\begin{figure}[tb!]
	\centering
		\subfigure[Initial Condition]{
			\begin{minipage}[b]{0.45\textwidth}
				\includegraphics[width=1\textwidth]{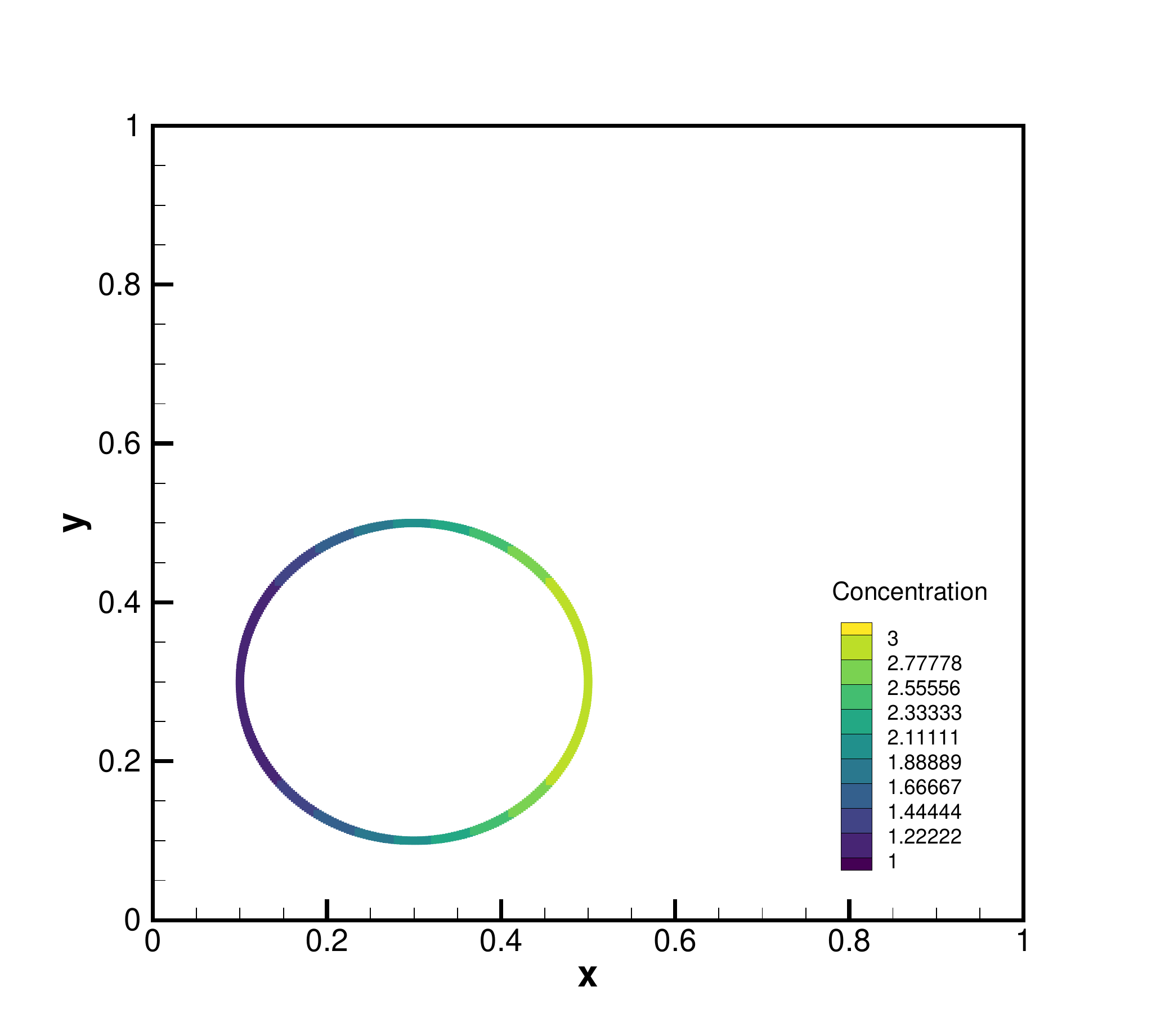}
			\end{minipage}
			\label{fig:TNI256}
		}
    	\subfigure[$t=1.0$]{
    		\begin{minipage}[b]{0.45\textwidth}
   		 	\includegraphics[width=1\textwidth]{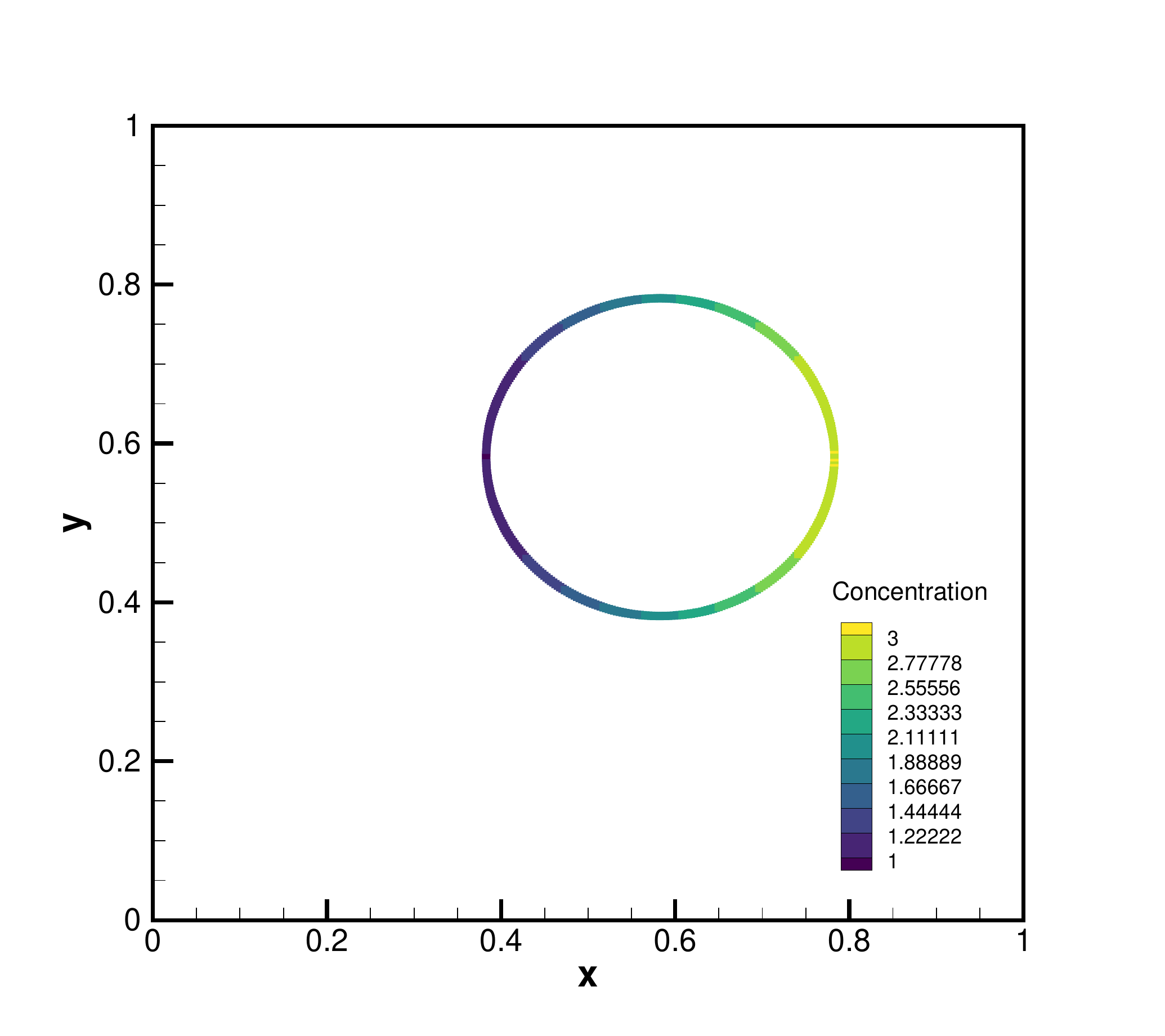}
    		\end{minipage}
			\label{fig:TNF256}
		}
    	
	\caption{Translation case with non-uniform initial concentration,$256\times256$ cells.}
	\label{fig:TN}
\end{figure}

	Nnumerical diffusion is tested by a non-uniform case.
	From the results in Fig. \ref{fig:TN} and Tab. \ref{non-uniform},
	we can see that the maximum and minimum concentration is preserved in the advection process,
	which implies the numerical diffusion is quite small.
\begin{table}[tb!]
	\caption{The relative error of translation case with uniform concentration}
\centering
	\begin{tabular}{|l|l|l|l|l|}
 		\hline
 		\quad Resolution & $64\times 64$ &$96\times 96$ & $128\times128$ & $256\times256$\\
 		\hline
 		Relative Error &  \ 1.932\% &1.005\% &  \quad 0.879\% & \quad 0.994\%\\
 		\hline
	\end{tabular}
	\label{uniformtab}
\end{table}
\begin{table}[tb!]
	\caption{The range of concentration in non-uniform case}
	\centering
	\begin{tabular}{|l|l|l|l|l|}
		\hline
		Resolution	&\quad$64\times 64$ & \ \quad$128\times128$ &\quad$256\times256$\\
		\hline	
		Range		&[0.999,\ 2.992]	&[1.0018,\ 3.00174]		   &[0.999,\ 3.002]\\
		\hline
	\end{tabular}
\label{non-uniform}
\end{table}
\subsection{Droplet in the simple shear flow}
\label{chap:simpleshear}
A circle in the simple shear flow can have a large deformation \cite{xu2003eulerian},
if there is no surface tension.
We compare the results with the analytical solution to verify the accuracy of our method.
\subsubsection{Derivation of the analytical solution}
The governing equation is the mass conservation:
\begin{equation}
\label{massconservation}
	\frac{d}{dt}\int_{L(t)}{C(l,t)dl} =0,
\end{equation}
where $L(t)$ is an arbitrary element on the interface at time $t$,
and $l$ is the arc length.
In other words, the total mass $M_L$ is constant on an element
\begin{equation}
	\label{conservation}
    \int_{L(t)}{C(l,t)dl} =M_L\bigg|_{(t=0)}.
\end{equation}	
The right hand side can be obtained by the initial condition.
The initial condition is that the interface is a unit circle centered at origin, and the passive scalar has a uniform distribution $C=1$, thus $M_L=\theta_1-\theta_0$, where $\theta_0$ and $\theta_1$ are the two ends of the considered curve $L(0)$. 
$L(t)$ is continuously deformed by the given velocity field, thus it can be determined by velocity field and initial element $L(0)$.
We assume that the flow is: $u=0$, $v=Mx$. L(0) can be parameterized as
\begin{align}
	x(\theta)&=\cos\theta,\\
	y(\theta)&=\sin\theta,
\end{align}
where $\theta \in [\theta_0,\theta_1]$.
Points on $L(t)$ are given as
\begin{align}
    x(\theta,t)&=\cos\theta,\label{x}\\
    y(\theta,t)&=\sin\theta+Mxt=\sin\theta+Mt\cos\theta,\label{y}
    \end{align}
and $dl$ can be represented by
\begin{equation}
	dl=\left|\frac{\partial }{\partial \theta}\vec{r}(\theta,t)\right|d\theta,
\end{equation}
where $\vec{r}(\theta,t)=(x(\theta,t),y(\theta,t))$.
Substituting this into Eq. (\ref{conservation}{}), we have
\begin{equation}
	\int_{\theta_0}^{\theta_1}{C(\theta,t)\left|\frac{\partial }{\partial \theta}\vec{r}(\theta,t)\right|d\theta} =\theta_1-\theta_0,
\end{equation}
which holds for arbitrary $\theta_0$ and $\theta_1$.
Consider the partial derivative with respect to $\theta_1$, we have
\begin{equation}
	C(\theta_1,t)\left|\frac{\partial }{\partial \theta}\vec{r}(\theta_1,t)\right|=1.
	\label{masscons}
\end{equation}
The second term in the left hand side can be obtained from Eq. (\ref{x}-\ref{y}). For this simple shear flow case, we have
\begin{equation}
	\left|\frac{\partial }{\partial \theta}\vec{r}(\theta,t)\right|=\sqrt{1+M^2t^2\sin^2\theta-2Mt\sin\theta \cos\theta}.
\end{equation}
Therefore, the solution for arbitrary $\theta$ is
\begin{equation}
	C(\theta,t)=\frac{1}{\sqrt{1+M^2t^2\sin^2\theta-2Mt\sin\theta \cos\theta}}.
	\end{equation}
Here we use $\theta$ to replace $\theta_1$ because $\theta_1$ can be arbitrary in Eq. (\ref{masscons}).
\subsubsection{Numerical Results}
\begin{figure}[tb!]
	\subfigure[Initial Condition]{
	\begin{minipage}[b]{0.45\textwidth}
		\includegraphics[width=1\textwidth]{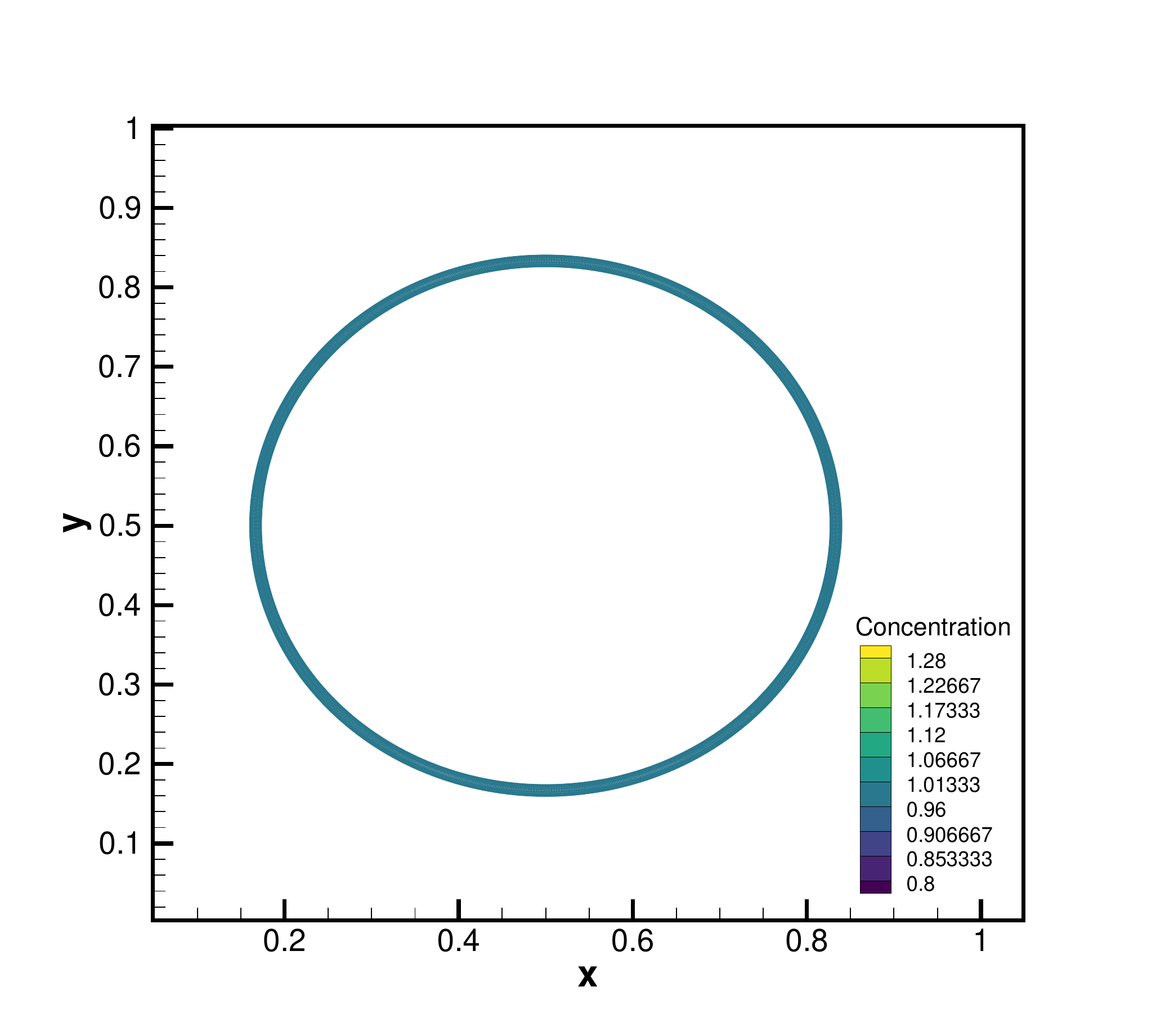}
	\end{minipage}
	}
    \subfigure[$M=0.5$, time$=1.0$]{
    	\begin{minipage}[b]{0.45\textwidth}
   		\includegraphics[width=1\textwidth]{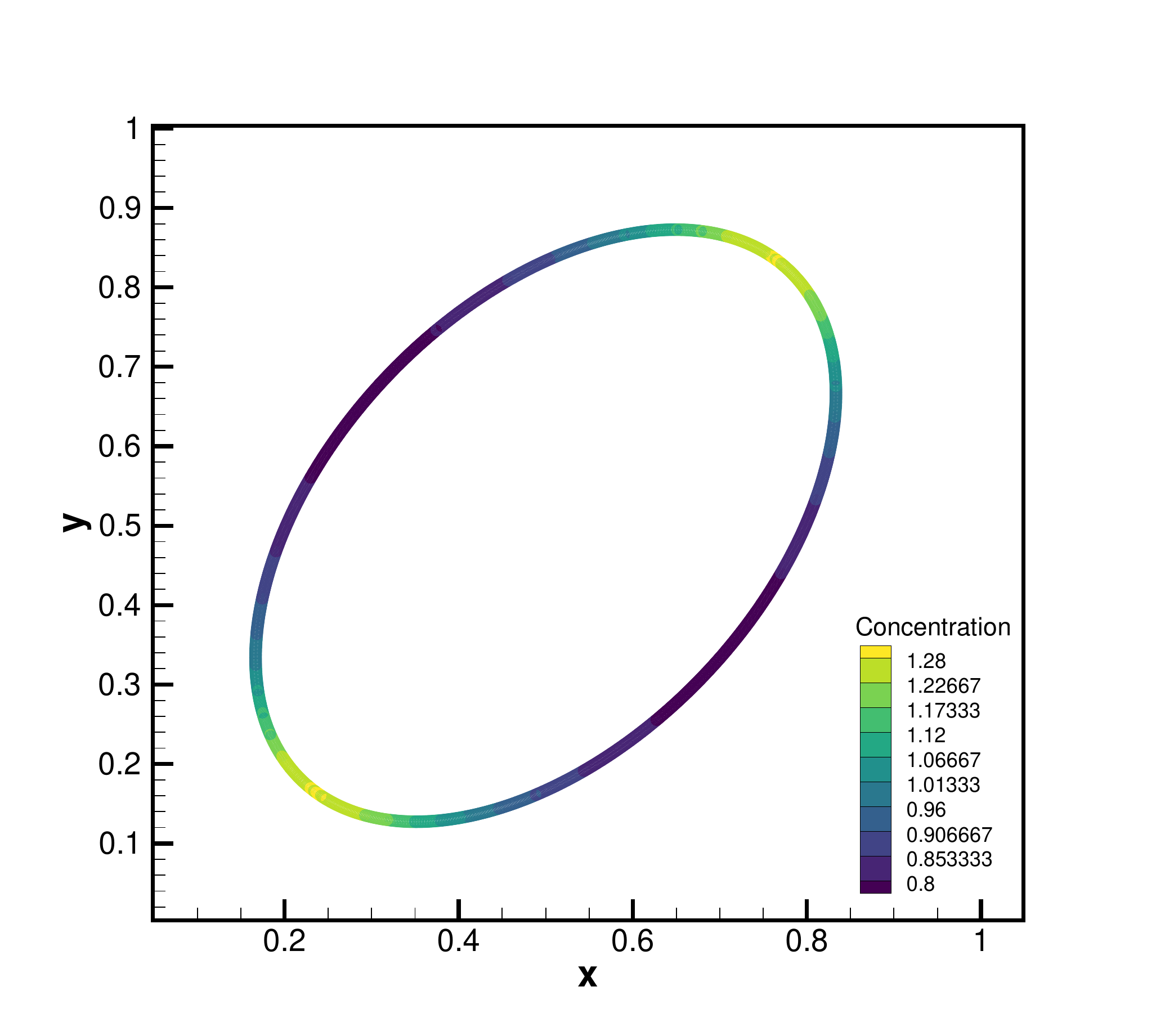}
    	\end{minipage}
	}
	\caption{A circle in simple shear flow}
	\label{SF}
\end{figure}
In this test case, $M=0.5$, and the final time $t=1.0$. The particles are remeshed after every step, with a target distribution quality $Q_0=Q_1=1\%$. The maximum value of the passive scalar exact solution is about $1.2806$. It can be shown from the Fig. \ref{SF} that the maximum value of the numerical result is about $1.28$. The $L_{\infty}$ norm of relative error of concentration field is shown in Tab. \ref{givensheartab} . 
	\begin{table}[tb!]
		\caption{The relative error of concentration field on droplet in the given shear flow.}
		\centering
		\begin{tabular}{|c|c|c|c|}
			\hline
			Resolution & $128\times 128$ &$256\times256$ &$512\times512$\\
			\hline
			Relative Error &0.504\% &  \ 0.397\% & \  0.289\%\\
			\hline
		\end{tabular}
		\label{givensheartab}
	\end{table}

We emphasize that, in addition to remeshing, errors in the simulation results may be introduced by other factors, such as the particle distribution and the consistency of the SPH method. Although sufficient iterations of relaxation have been done, the particle distribution is not perfectly uniform. Even with a perfectly uniform one, the SPH method still can not achieve the zero-th order exactly. 
Thus, these errors cannot be remedied solely by increasing the resolution of the simulation.

\subsection{Drop deformation Case}
We consider the deformation of a droplet immersed in another shearing fluid \cite{taylor1932viscosity} to verify our method in two-phase flow.
There are two walls at top and bottom of the bulk fluid moving in opposite directions with constant velocity $\pm u_\infty$,
which induces a shear flow in the bulk phase, 
and finally leads to the deformation of the immersed droplet.
The deformation was measured experimentally \cite{taylor1934formation}, theoretically \cite{taylor1932viscosity} 
and by numerical simulation \cite{adami2010conservative,luo2015conservative,hu2007incompressible}.
The geometry is $8R_0 \times 8R_0$ in $XoZ$ plane,
and the center of the drop is $(4R_0,4R_0)$.
The ratio of the viscosity $\lambda = \eta_d/\eta_w=100$.
The shear rate is
$G=2u_\infty/8R_0=1$.
The non-dimensionless parameters are
\begin{align}
	&Ca = \frac{G\eta_w R_0}{\alpha}=0.15,\\
	&Re = \frac{\rho G R_0^2}{\eta_w}=1.
\end{align}
We use the fluid solver in \cite{luo2015conservative}, which has been verified before.
The different thing is that
we consider the variations of the concentration field
on the interface with an uniform initial distribution $C=1$.

\begin{figure}[tb!]
	\centering
	\includegraphics[width=0.8\textwidth]{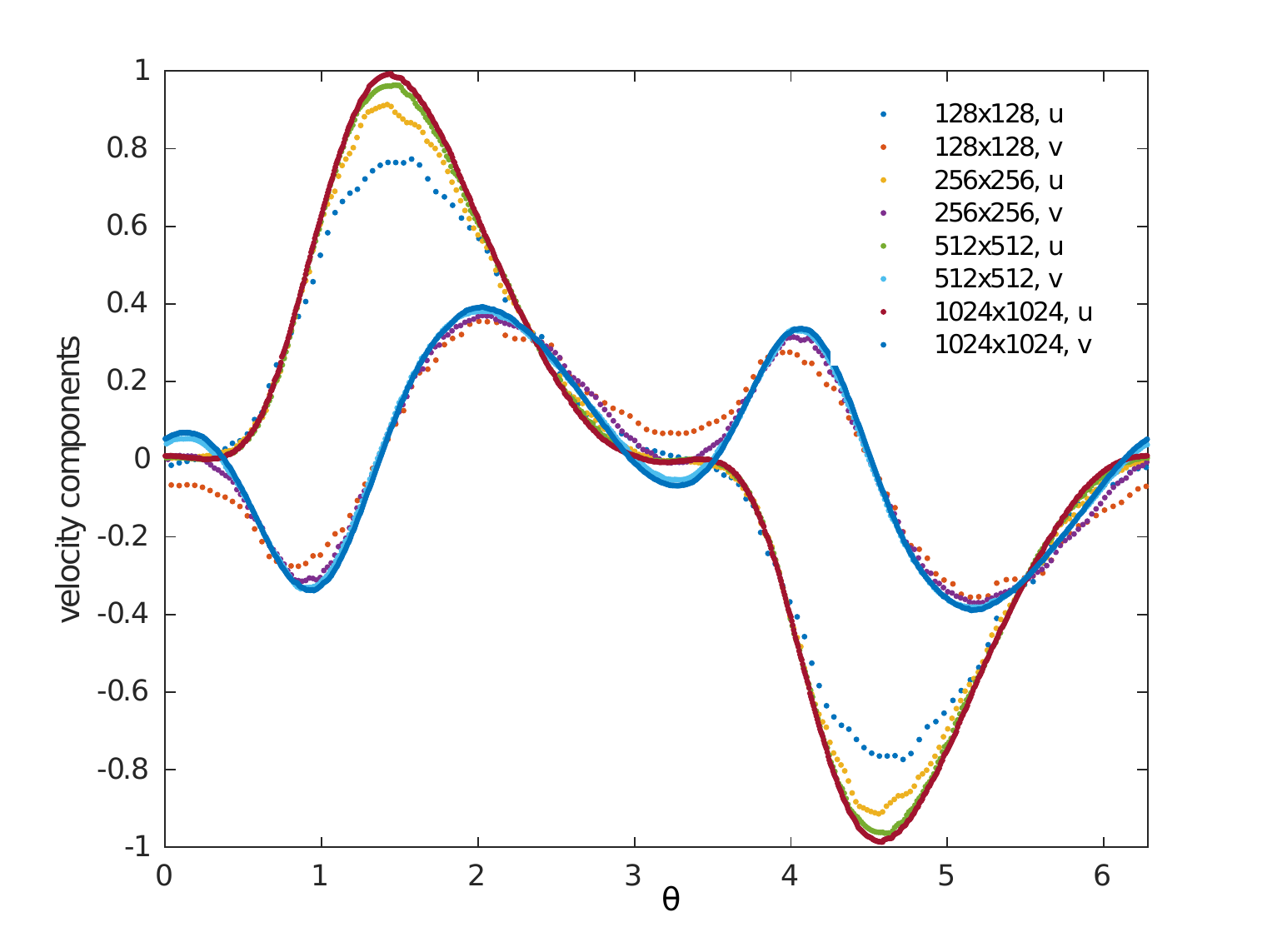}
	\caption{The convergence study of interfacial velocity components}
	\label{components_vel}
\end{figure}

In the past study, the convergence of the deformation is studied, and the shape almost converges even in a low resolution, like $128\times 128$.
If only the interface advection is considered, the surface concentration is only dependent on the interface velocity.
We sample the interfacial velocity on the particles and find that
the convergence for the velocity requires much higher resolution than the shape, see Fig. \ref{components_vel}.
\begin{figure}[tb!]
	\centering
	\includegraphics[width=0.8\textwidth]{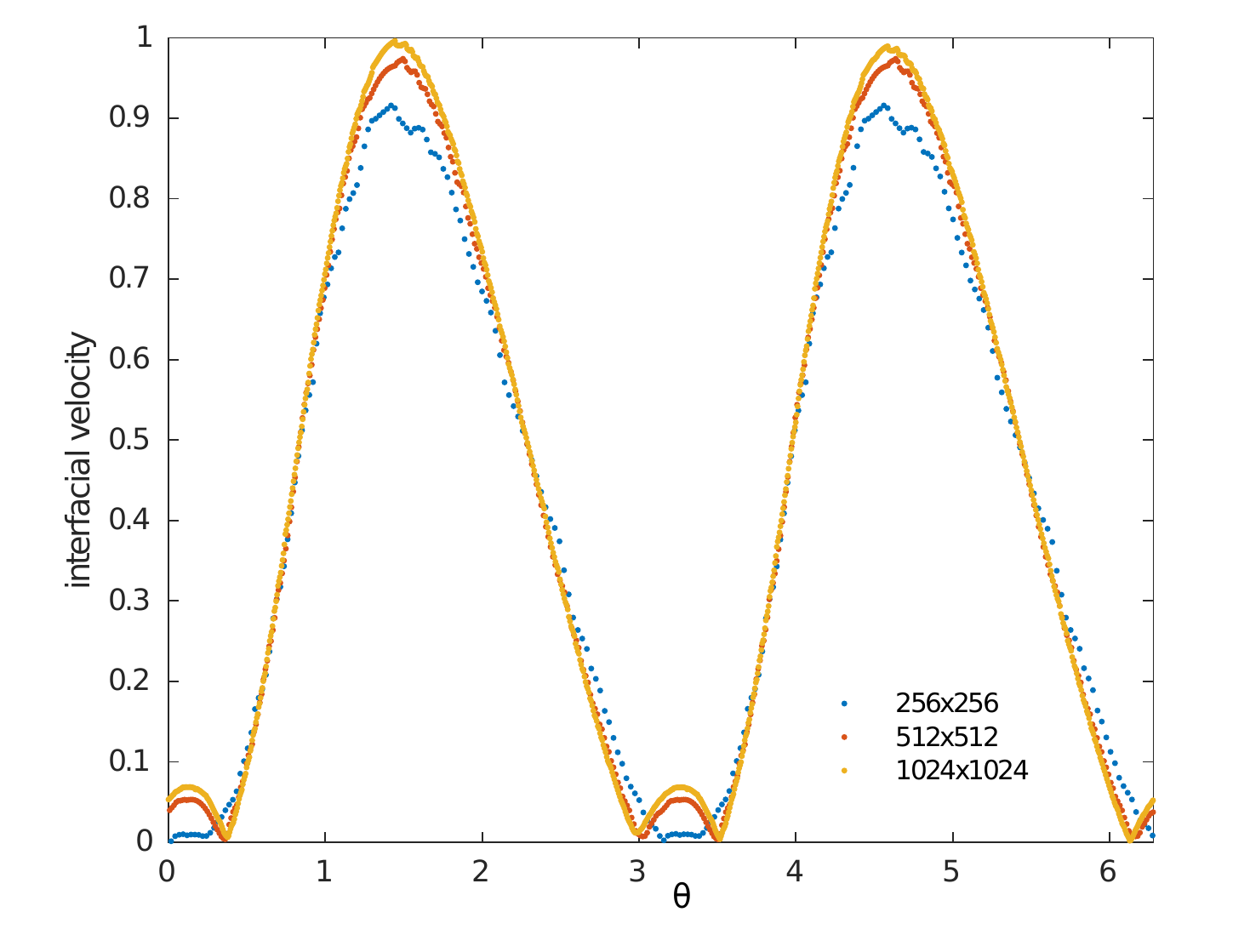}
	\caption{The convergence study of interfacial velocity magnitude}
	\label{convergence_vel}
\end{figure}
Finally we observe from Fig. \ref{convergence_vel} that the velocity almost converges above the $1024\times 1024$ resolution,
but usually $128 \times 128$ or $256 \times 256$ (or equivalent number of particles in particle methods) is enough to track the interface deformation \cite{hu2006conservative,hu2007incompressible}.
It is worth stressing that the convergence of velocity is just the prerequisite for the convergence of the concentration for pure advection.
We can see that the magnitude of the interfacial velocity is highly non-uniform, which results in some points being highly compressed while others are stretched. This non-uniformity makes convergence more difficult. If there is no diffusion, the concentration near the compressed region would become infinitely high, which is unphysical and violates the mono-layer assumption. Therefore, a mechanism is needed to constrain the concentration from approaching infinity.
Therefore, we introduce a diffusion term into the passive scalar transport and the transport equation becomes
\begin{equation}
	\frac{\partial C}{\partial t}+C\nabla_{s} \cdot \vec{u}=D\nabla^2_{s}C.
\end{equation}
The discretized scheme is
\begin{equation}
	\frac{Dm}{Dt}\bigg|_i=2D\sum_{k \neq i}{\frac{(C_i-C_k)}{r_{ik}}}W'_{ik}V_iV_k,
	\label{diffequ}
\end{equation}
where the $D = 1.0$ is the diffusion coefficient and the corresponding Peclet number is $Pe =GR_0^2/D=1$. 
Eq. (\ref{diffequ}) is based on \cite{brookshaw1985method}.
We modify it from the concentration to the mass by multiplying the volume of the particle.

\begin{figure}[tb!]
    \subfigure[Initial Condition, $1024\times 1024$]{
		\begin{minipage}[b]{0.45\textwidth}
			\includegraphics[width=1\textwidth]{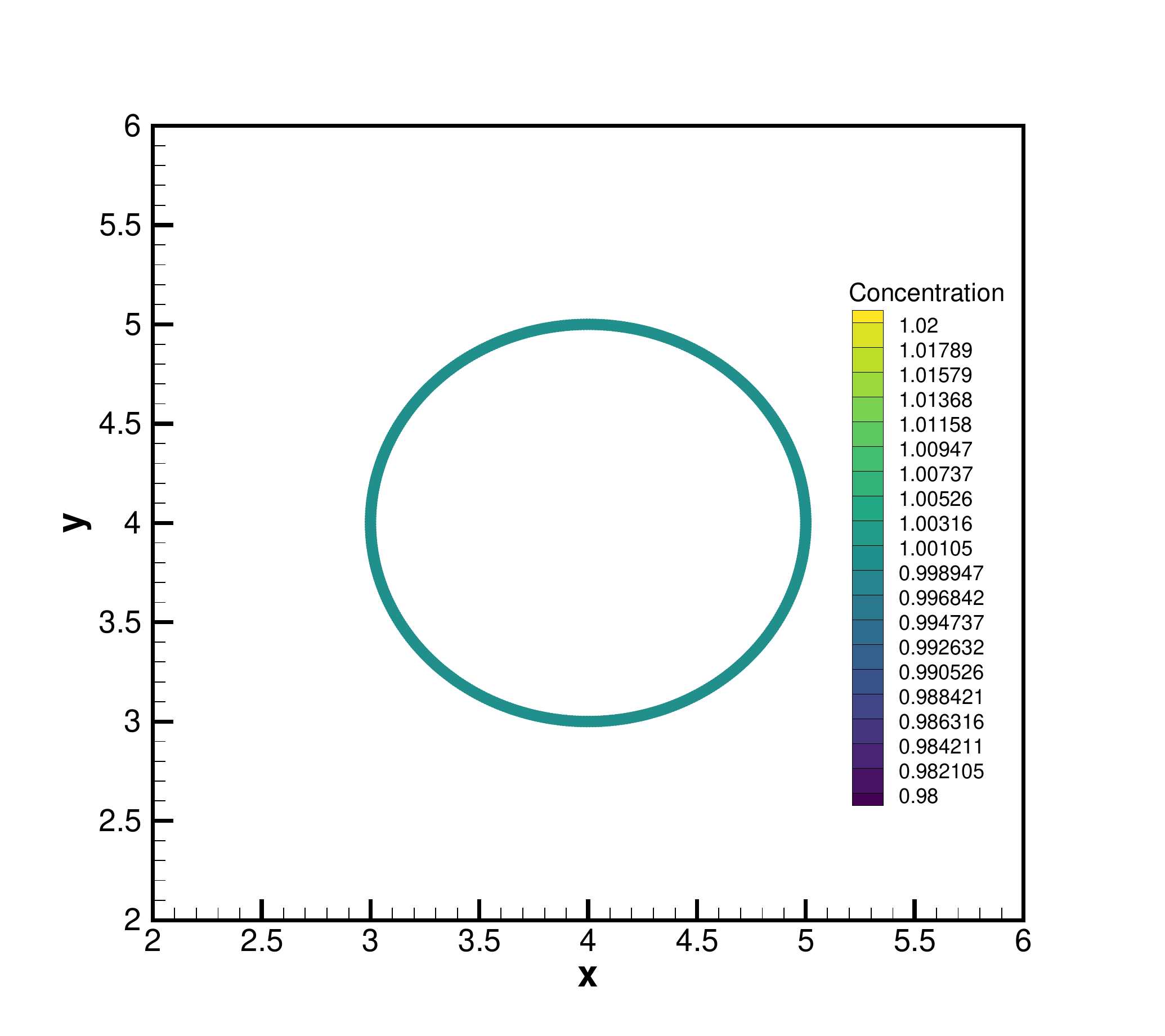}
		\end{minipage}
	}
	\subfigure[$T=8,\ 256\times 256$, $D = 0.1886$]{
		\begin{minipage}[b]{0.45\textwidth}
			\includegraphics[width=1\textwidth]{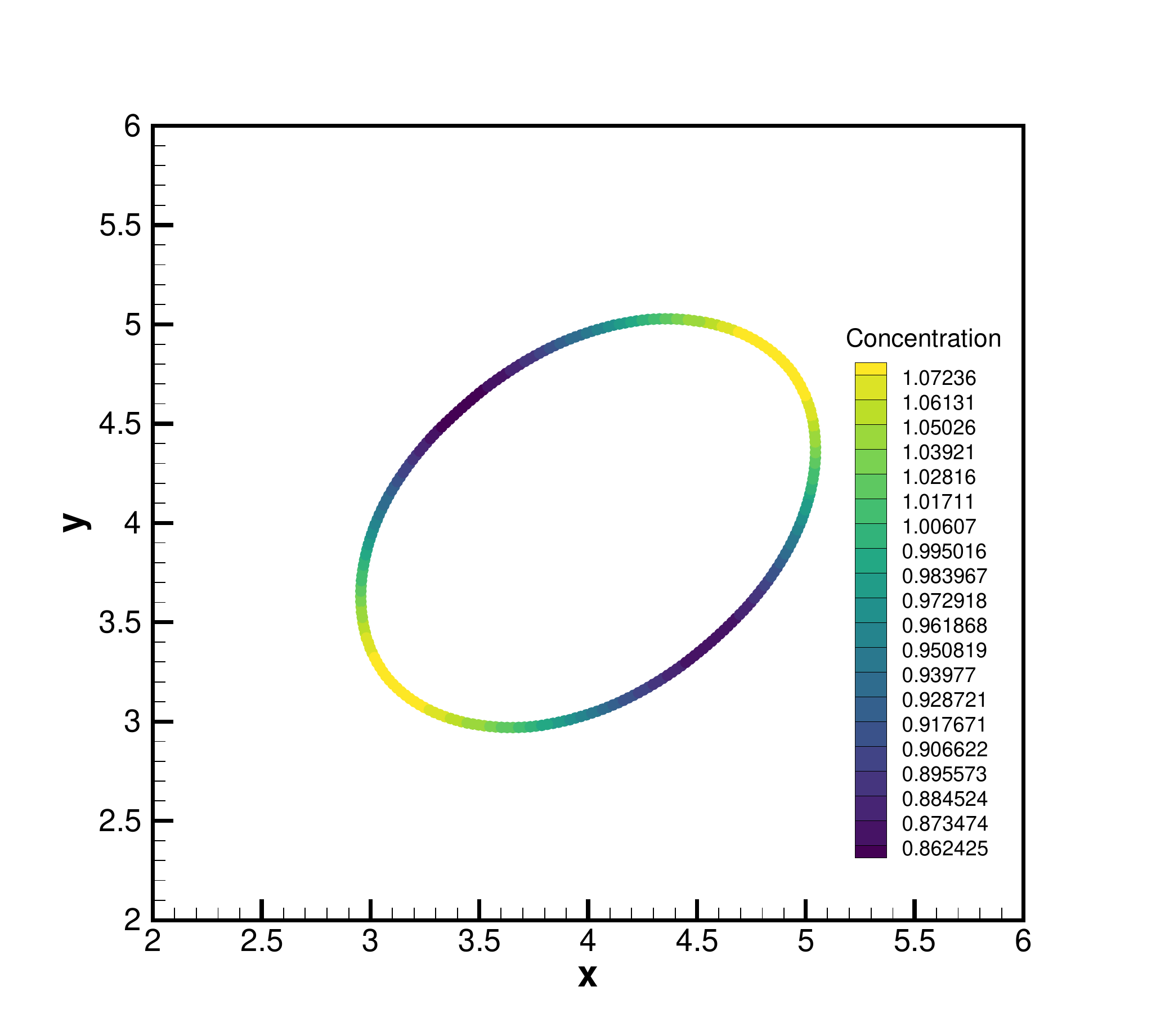}
		\end{minipage}
	}
		\\
   	\subfigure[$T=8,\ 512\times 512$, $D = 0.2069$]{
   		\begin{minipage}[b]{0.45\textwidth}
  		 	\includegraphics[width=1\textwidth]{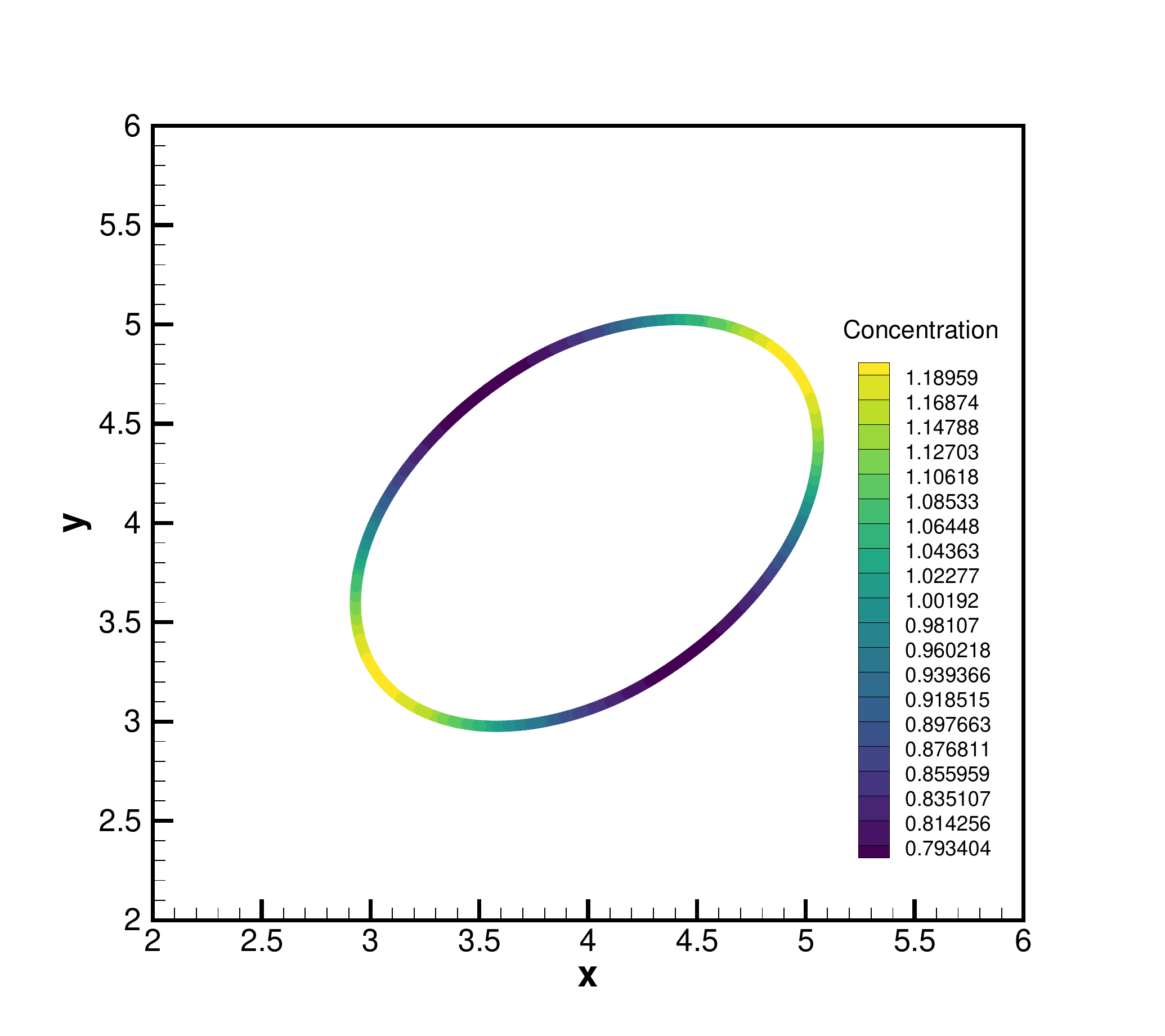}
   		\end{minipage}
	}
   	\subfigure[$T=8,\ 1024\times 1024$, $D=0.2102$]{
   		\begin{minipage}[b]{0.45\textwidth}
  		 	\includegraphics[width=1\textwidth]{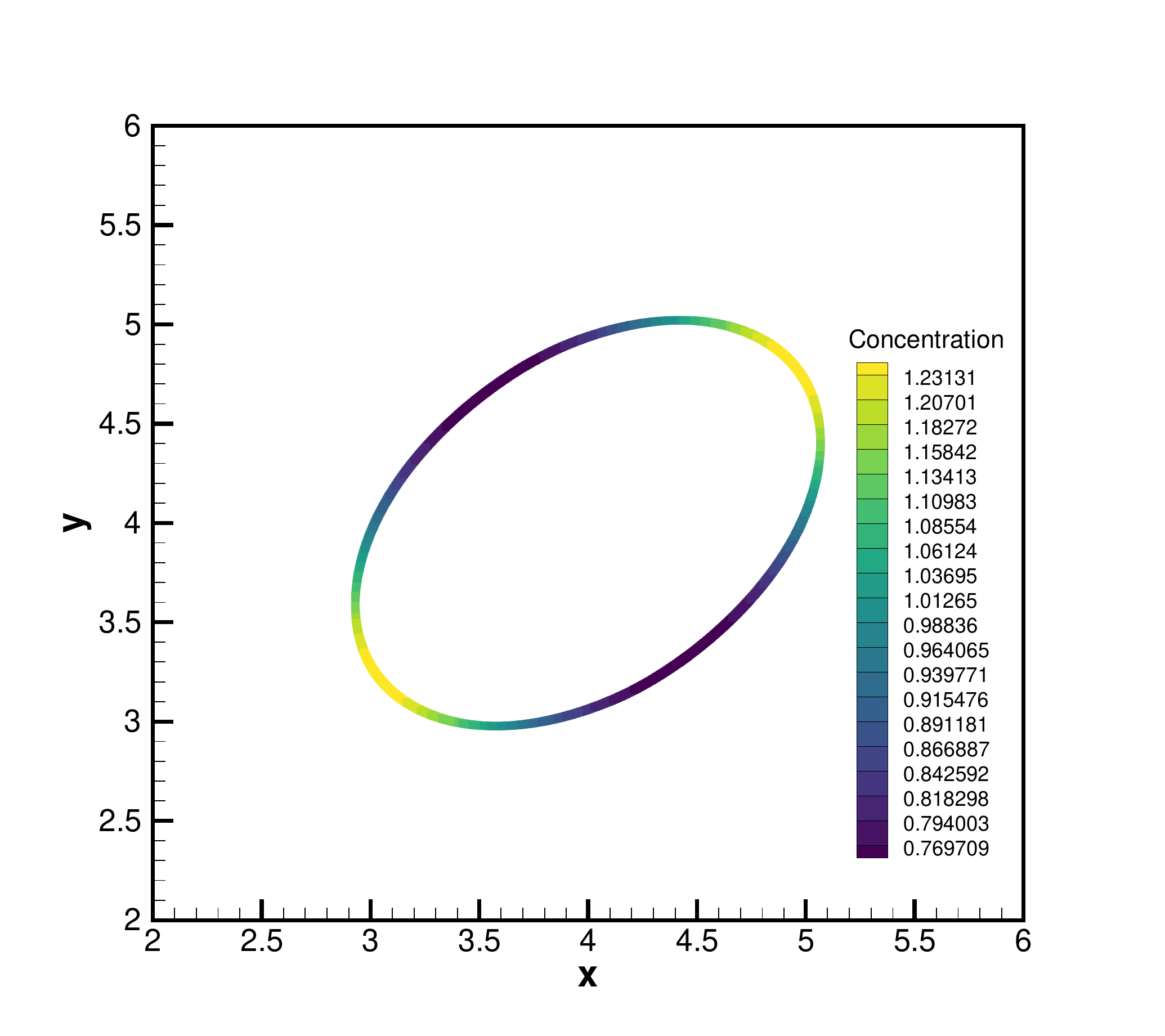}
   		\end{minipage}
	}
	\caption{A droplet in shear flow with different resolution, $Q_0 = 1\%$ and $Q_1 = 3\%$.}
	\label{DSF}
\end{figure}

After introducing the diffusion term, the convergence of the concentration field is improved since the diffusion term helps to smooth out large concentrations caused by high compression, see Fig. \ref{DSF}. Moreover, the deformation can be measured by a deformation parameter which is defined as
\begin{equation}
D = Ca \frac{16+19\lambda}{16+16\lambda}.
\end{equation}
The theoretical and numerical deformation parameters for a clean droplet are 0.1778 and 0.1718 (with resolution $512 \times 512$).
From Fig. \ref{DSF}, we can see that the deformation parameter is almost convergent at resolution of $1024\times 1024$, and the parameter is much larger than the clean one, which implies the Marangoni force plays an important role.

We would like to stress that relaxation iterations are usually very expensive. However, with our method, relaxation is only invoked when the particle distribution is not sufficiently uniform. Moreover, with the adaptive strategy, overheads are drastically reduced. We measured the costs (wall clock time) in a $128 \times 128$ droplet deformation case on a workstation with an I9-10980XE 18-core CPU. The relaxation process only costs a total of $0.959$ seconds, which is a very small fraction compared to the full cost of $2362.11$ seconds.

\subsection{Multi-region case: Two tracks case}
\begin{figure}[tb!]
    \subfigure[Uniform initial condition]{
		\begin{minipage}[b]{0.45\textwidth}
			\includegraphics[width=1\textwidth]{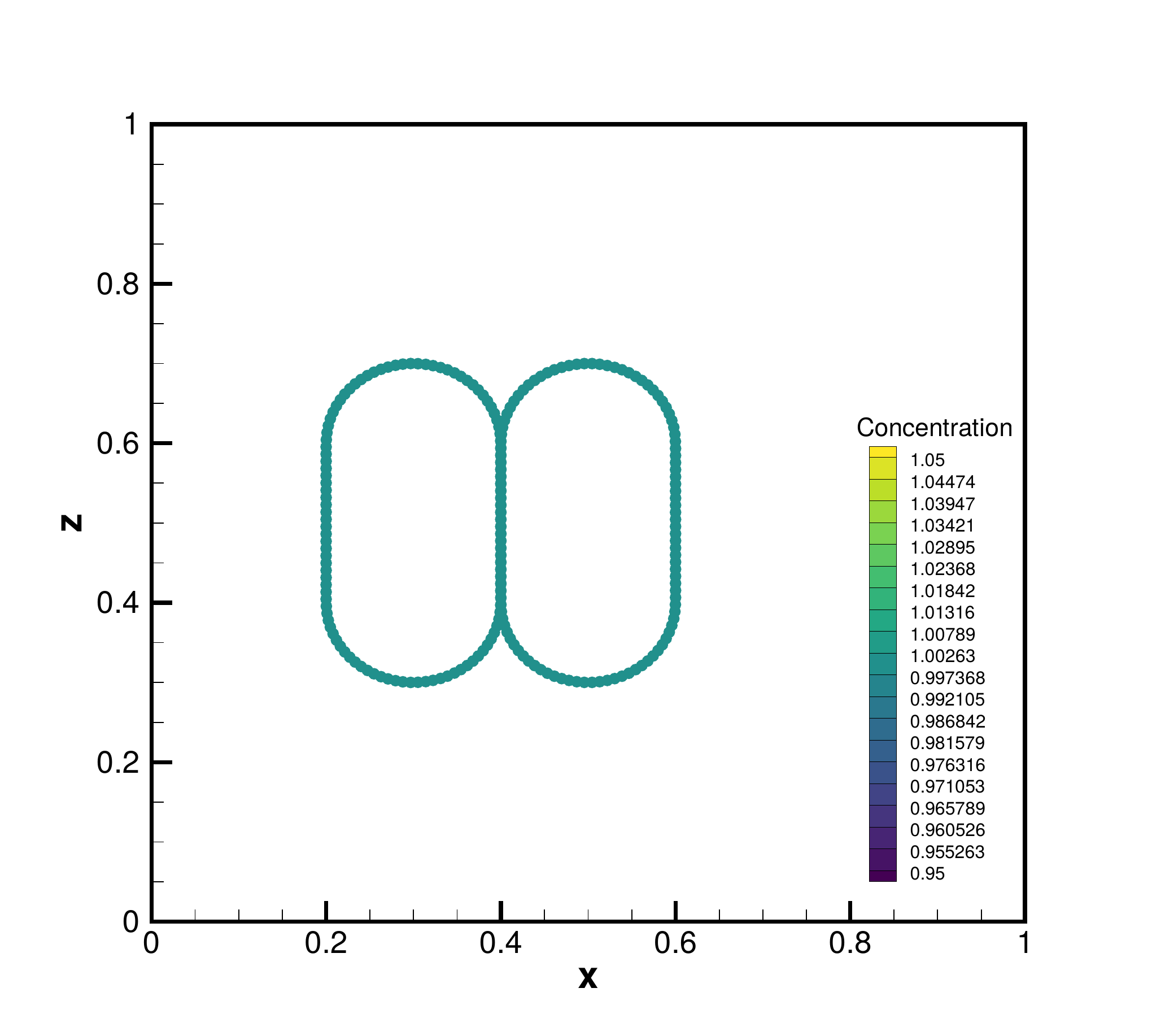}
		\end{minipage}
		\label{multi-shape}
	}
    \subfigure[Uniform case, time$=1.0$]{
    	\begin{minipage}[b]{0.45\textwidth}
   		 	\includegraphics[width=1\textwidth]{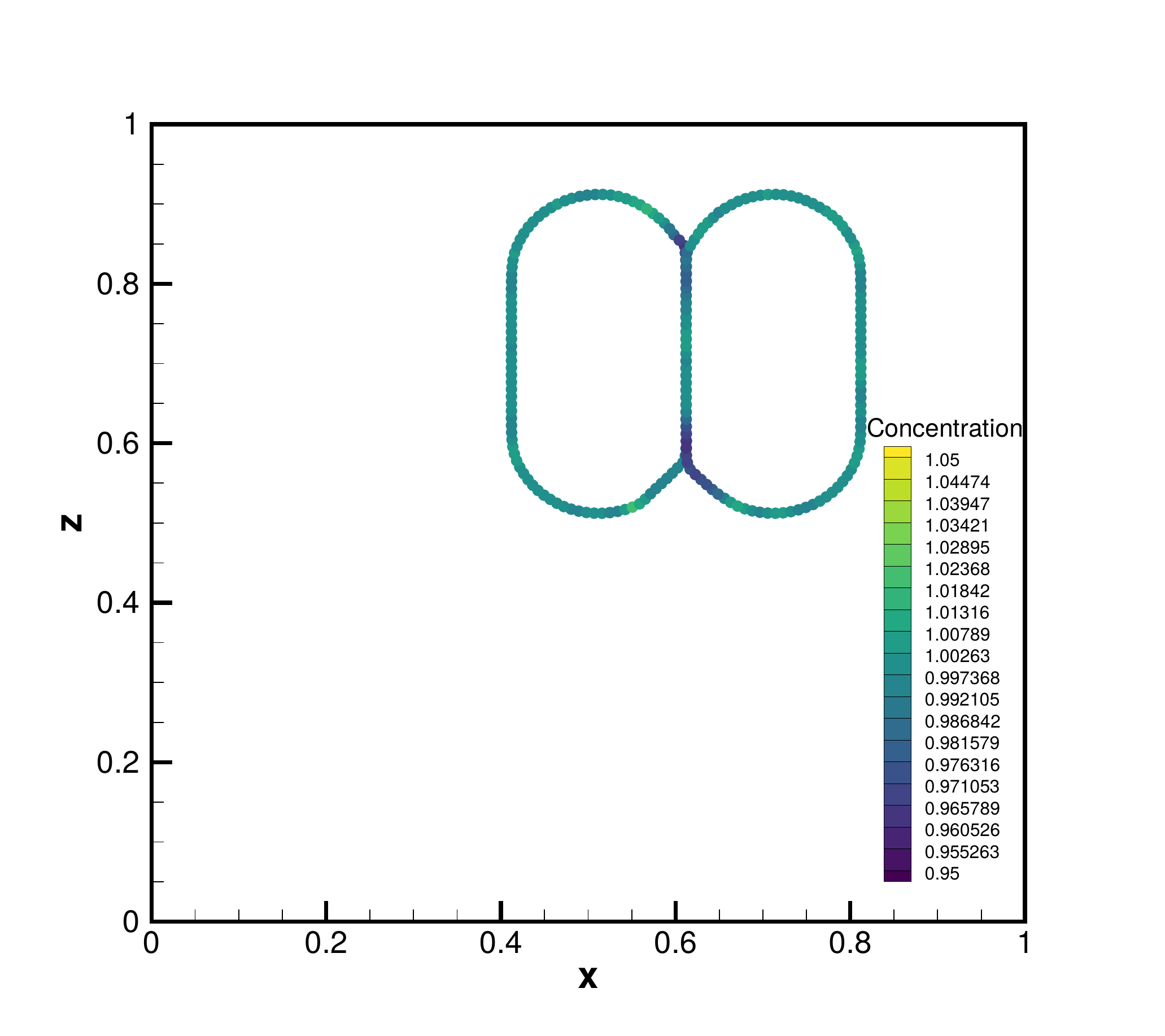}
    	\end{minipage}
	}
	\\
	 \subfigure[Non-uniform initial condition]{
		\begin{minipage}[b]{0.45\textwidth}
			\includegraphics[width=1\textwidth]{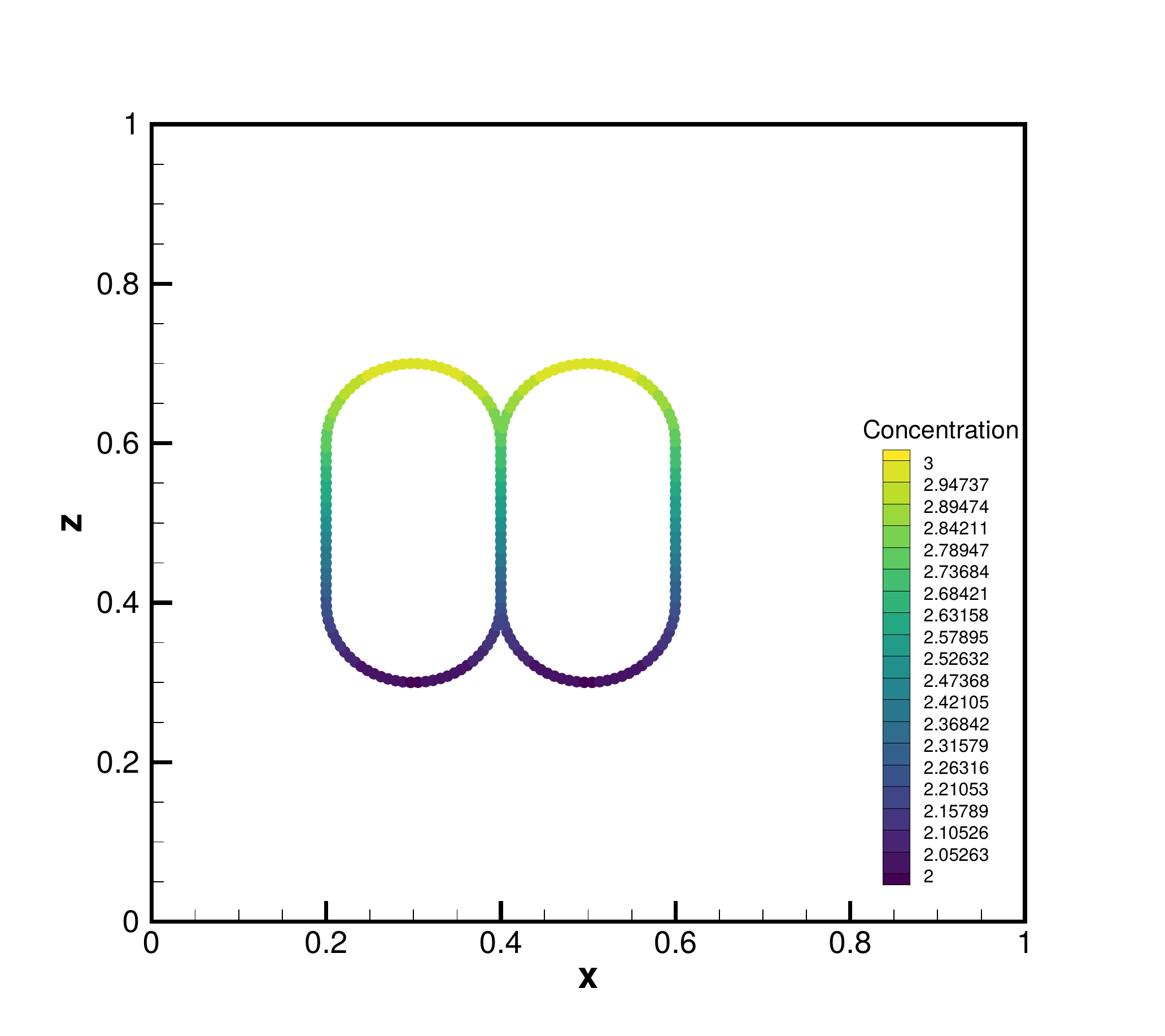}
		\end{minipage}
	}
    \subfigure[non-uniform, time$=1.0$]{
    	\begin{minipage}[b]{0.45\textwidth}
   		 	\includegraphics[width=1\textwidth]{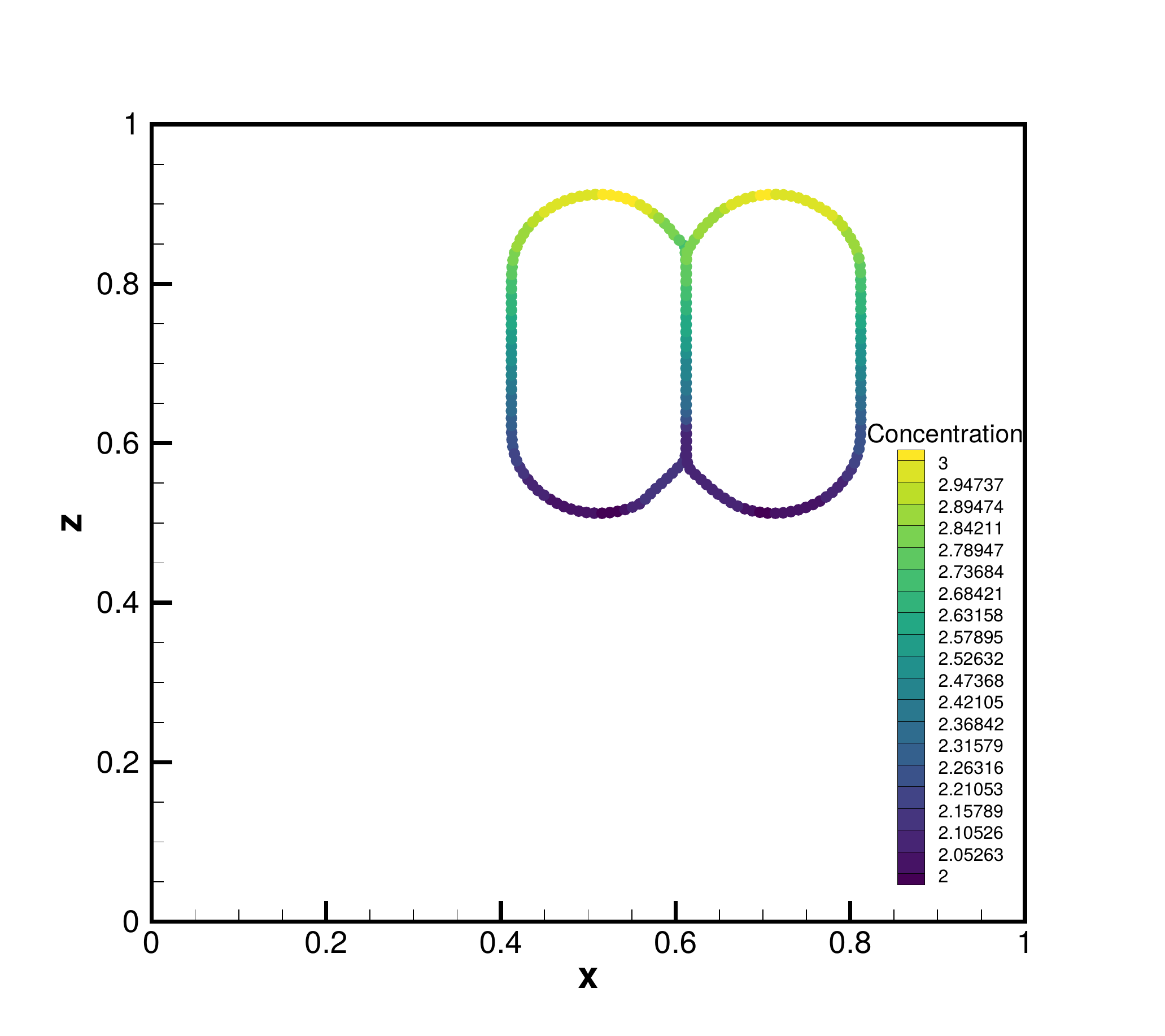}
    	\end{minipage}
	}
	\caption{Translation of two tracks}
	\label{Track128}
\end{figure}

This case is constructed by the following steps.
First, on the $(x,z)$ plane, we draw two circles with $r=0.1$ close to each other,
centered at $x_1=0.3$, $x_2=0.5$.
We slice the two circles at $z=0.5$.
Then we insert straight lines in between the $4$ half circles with length 0.2, see Fig. (\ref{multi-shape}).

The computational domain is $1.0 \times 1.0$. The background velocity is given as $u=w=0.3/\sqrt{2} $. The resolution in this case is $128\times128$.
For the uniform case, the initial condition is $C = 1$. For the non-uniform case, the initial concentration depends linearly on the height from $2$ to $3$. 
Such a multi-region case will not produce the singularity,
namely the interface near the triple point is $C^1$ smooth.

\begin{figure}[tb!]
    \subfigure[Interface for left region, time$=1.0$]{
		\begin{minipage}[b]{0.45\textwidth}
			\includegraphics[width=1\textwidth]{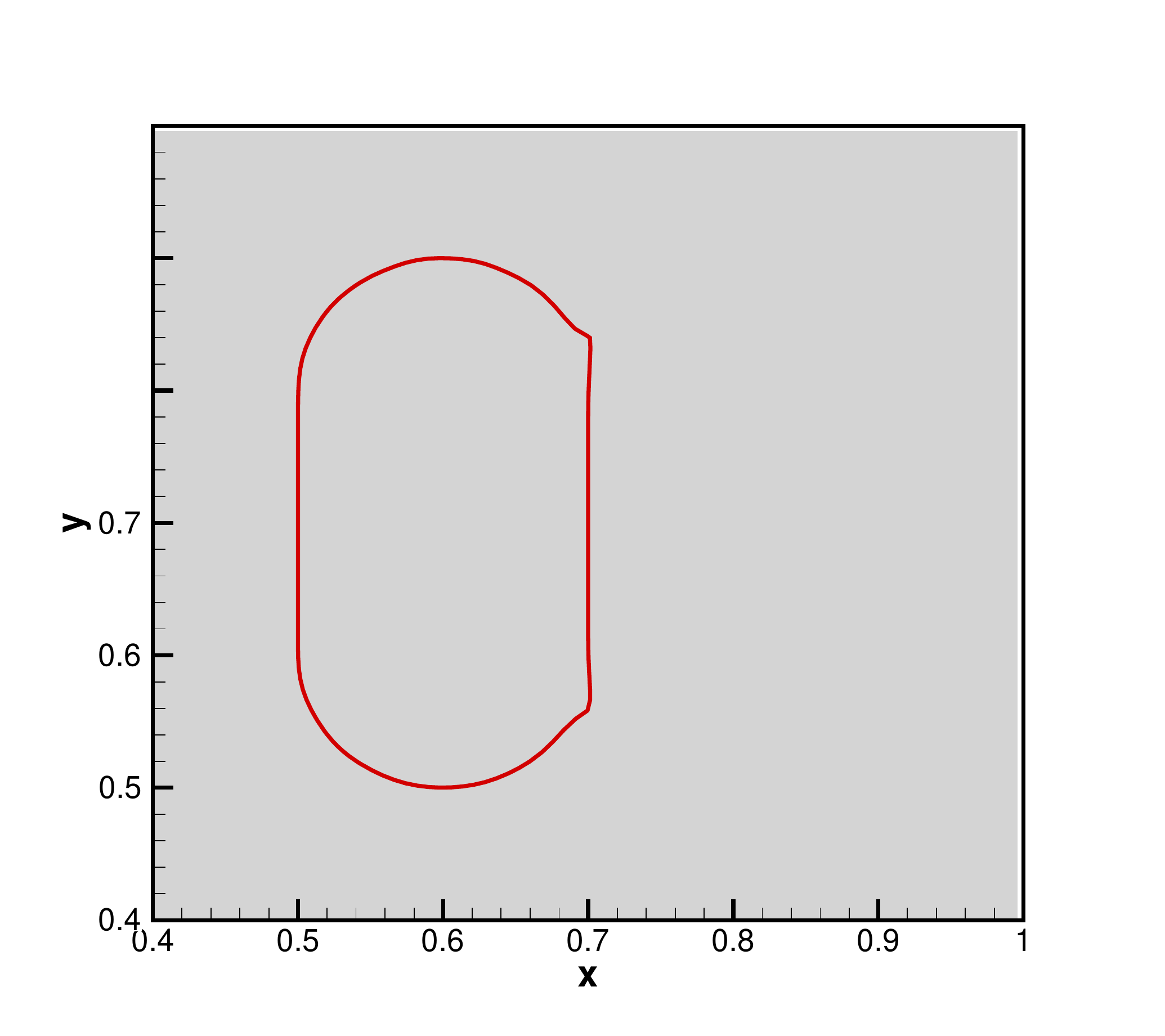}
		\end{minipage}
	}
    \subfigure[Concentration distribution, about $16\%$  error, time$=1.0$]{
    	\begin{minipage}[b]{0.45\textwidth}
   		 	\includegraphics[width=1\textwidth]{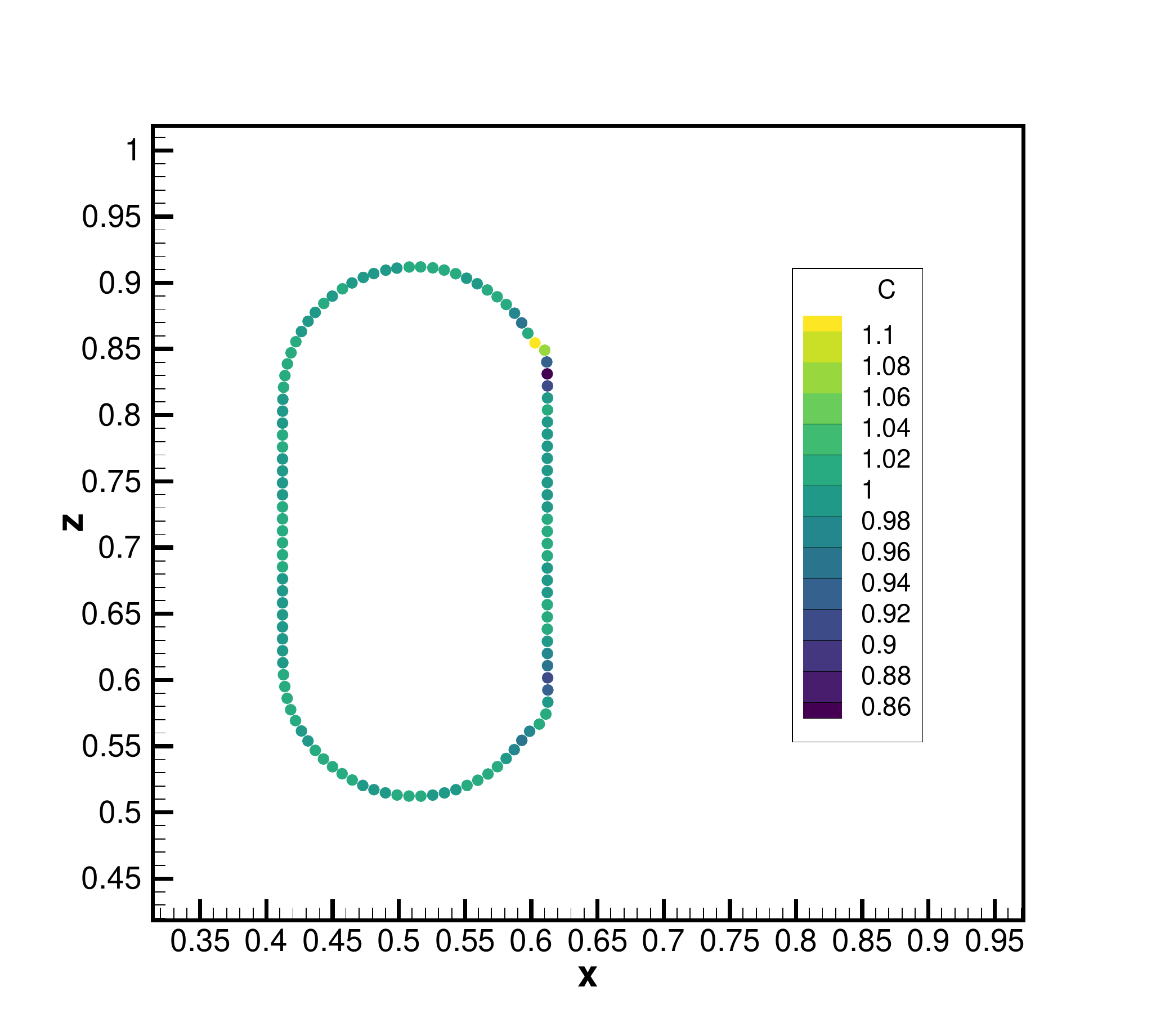}
    	\end{minipage}
	}
	\caption{The results for the left region, without modifications for uniform initial concentration.}
	\label{shucheng}
\end{figure}

From Fig. \ref{Track128},
we can observe that the relative error of concentration near the triple point is approximately $3.5\%$ in the uniform case. This error is relatively higher compared to other cases due to the less accurate regional level set near the triple point, despite our efforts to enhance its accuracy through various techniques in Sec. \ref{correctionsrl}. However, these techniques have significantly improved the accuracy, resulting in a four to five times improvement as illustrated in Fig. \ref{shucheng}.
If the local level-set fields near triple points can be further improved, both the accuracy and convergence reported in Section \ref{chap:translation} and Section \ref{chap:simpleshear} can be achieved.

\section{Conclusion}
To simulate the concentration evolution of a passive scalar on a moving interface,
we propose a hybrid Eulerian-Lagrangian method inherits the advantages of either side.
Conservation of mass is preserved by the particle method,
and the level-set method captures the evolution of the interface.
For validation, we use a circle with a translational motion to prove the accuracy.
The error introduced by SPH method and generalized remeshing method is reasonably small
and can be smaller as resolution increases.
Subsequently, we test a droplet in a given shear flow with equally good results.
Furthermore, the droplet in the two-phase shear flow is validated with converging results.
Finally, the two-track case is devised and tested for multi-region problems.
Although the relaxation algorithm is expensive,
the overhead is well controlled by using interfacial particles with co-dimension one
and reducing the frequency of remeshing adaptively.
The Lagrangian interpolation, due to its unavailability in higher dimensions, presents a challenge in extending it to three-dimensional space, and requires further investigation in the future. Furthermore, the enhancement of regional level-set method accuracy near triple points remains an unresolved issue that warrants scholarly attention.

\section{References}
\bibliography{mybibfile}

\end{document}